\documentclass[12pt,draftclsnofoot,onecolumn]{IEEEtran}
\usepackage{graphicx}
\usepackage{subfigure}
\usepackage{amsmath,amsthm,amsfonts}
\usepackage{cite}
\usepackage{bm}
\usepackage{url}
\usepackage{array}
\usepackage{epstopdf}
\usepackage[colorlinks,linkcolor=black,anchorcolor=blue,citecolor=black]{hyperref}
\usepackage{enumerate}
\usepackage{mdwlist}
\usepackage{amsthm}

\allowdisplaybreaks[4]
\theoremstyle{remark}

\begin{document}

\title{Noncoherent Detection for Physical-Layer Network Coding}
\author{Zhaorui~Wang,
        Soung~Chang~Liew,~\IEEEmembership{Fellow,~IEEE,}
        and~Lu~Lu,~\IEEEmembership{Member,~IEEE}
\thanks{Z. Wang and S. C. Liew are with the Department of Information Engineering, The Chinese University of Hong Kong, Shatin, New Territories, Hong Kong. Email:~\{wz015,~soung\}@ie.cuhk.edu.hk. L. Lu is with the Technology and Engineering Center for Space Utilization, Chinese Academy of Sciences, Beijing 100864, China. He was with the Institute of Network Coding, The Chinese University of Hong Kong, Hong Kong. Email:~lulu@csu.ac.cn}
}
\maketitle
\markboth{Technical Report}{}%

\begin{abstract}
This paper investigates noncoherent detection in a two-way relay channel operated with physical layer network coding (PNC), assuming FSK modulation and \emph{short-packet} transmissions. For noncoherent detection, the detector has access to the magnitude but not the phase of the received signal. For conventional communication in which a receiver receives the signal from a transmitter only, the phase does not affect the magnitude, hence the performance of the noncoherent detector is independent of the phase. PNC, on the other hand, is a multiuser system in which a receiver receives signals from multiple transmitters simultaneously. The \emph{relative phase} of the signals from different transmitters affects the received signal magnitude through constructive-destructive interference. In particular, for good performance, the noncoherent detector of a multiuser system such as PNC must take into account the influence of the relative phase on the signal magnitude. Building on this observation, this paper delves into the fundamentals of PNC noncoherent detector design. To avoid excessive overhead, we assume a set-up in which the short packets in the PNC system do not have preambles. We show how the relative phase can be deduced directly from the magnitudes of the received data symbols, and that the knowledge of the relative phase thus deduced can in turn be used to enhance performance of noncoherent detection. Our overall detector design consists of two components: 1) a channel gains estimator that estimates channel gains without preambles; 2) a detector that builds on top of the estimated channel gains to jointly estimate relative phase and detect data using a belief propagation algorithm. Numerical results show that our detector performs nearly as well as a ``fictitious'' optimal detector that has perfect knowledge of the channel gains and relative phase. Although this paper focuses on PNC with FSK  modulation, we believe the insight of this paper applies generally to noncoherent detection in other multiuser systems with other modulations. Specifically, our insight is that the relative phase of overlapped signals affects the signal magnitude in multiuser systems, but fortunately the relative phase can be deduced from the magnitudes and this knowledge can be used to improve detection performance.
\end{abstract}


\section{Introduction}
\begin{figure}[ht]
  \centering
  \includegraphics[scale=0.80]{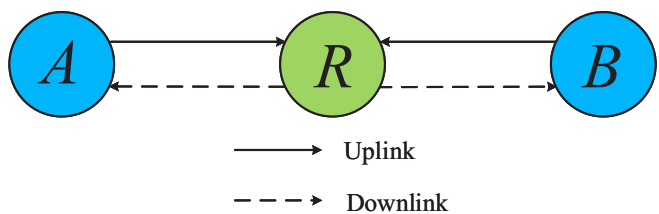}\\
  \caption{A two-way relay channel operated with physical-layer network coding (PNC), where two users \emph{A} and \emph{B} exchange messages via relay \emph{R}.}\label{Fig1}
\end{figure}

We study a two-way relay channel operated with physical layer network coding (PNC)\cite{zhang2006hot}\cite{popovski2007physical}\cite{nazer2011compute}, assuming \emph{frequency shift keying} (FSK) modulation\cite{goldsmith2005wireless} and \emph{short-packet} transmissions\cite{wang2014cellular}. Fig.\ref{Fig1} shows the model under consideration. Users \emph{A} and \emph{B} are out of each other's transmission range, and they exchange messages with the assistance of relay \emph{R}. There are two phases in the message-exchange mechanism. In the uplink phase, users \emph{A} and \emph{B} transmit their messages simultaneously to relay \emph{R}. From the overlapped signals, relay \emph{R} deduces a network-coded message. In the downlink phase, relay \emph{R} broadcasts the network-coded message to both users. User \emph{A} then uses the network-coded message and its own message to deduce message from user \emph{B}\cite{liew2013physical}. Likewise for user \emph{B}. An application scenario is message exchange among \emph{Internet of things} (IoT) that generate tiny messages. In this scenario, users \emph{A} and \emph{B} are IoT devices that exchange messages.

FSK encodes bit information into transmitted frequencies. For binary FSK, bit 0 corresponds to one frequency and bit 1 corresponds to another frequency. This paper investigates binary FSK in PNC (FSK-PNC). When users A and B both transmit bit 0 or bit 1, then their transmitted frequencies overlap; otherwise their transmitted frequencies are distinct.

We assume local oscillators (LOs) at \emph{A} and \emph{B} are low-cost and thus the frequencies generated from their LOs may not be highly accurate and stable. This means that the symbols of users \emph{A} and \emph{B} may have varying relative phase from symbol to symbol because the frequencies of their LOs may be slightly different and may drift in a different way. For simple circuit implementation and robust performance against phase variations, we consider the use of noncoherent detection\cite{goldsmith2005wireless}\cite{ProakisJohnG2008Dc}\cite{CouchLeonW2013Daac} at the receiver (relay \emph{R}) for the detection of PNC packets.

At the receiver, the noncoherent detector has access to the magnitudes of the received signals, but not the phase associated with the received signals\cite{goldsmith2005wireless}\cite{ProakisJohnG2008Dc}\cite{CouchLeonW2013Daac}. The magnitudes, for example, can be obtained by a simple signal-envelope detector\cite{CouchLeonW2013Daac}. In conventional single-user point-to-point communications where a transmitter transmits to a receiver, the magnitude and the phase of the received signal are independent. Therefore, the performance of the conventional noncoherent detector is not affected by the phase.

PNC, however, has two users transmitting signals simultaneously to a common receiver. In FSK-PNC, when the two users transmit on the same frequency, the relative phase between the two users will cause constructive-destructive interference. In particular, the magnitude of the superimposed signal at the receiver depends on the relative phase between two users. Because of that, performance of the noncoherent detector depends on the ``hidden'' relative phase. Due to this subtlety, noncoherent detection in PNC calls for an approach fundamentally different from noncoherent detection in conventional single-user communication systems. In particular, if the ``hidden'' relative phase can be uncovered from the available signal magnitudes, then the performance of the noncoherent detection in PNC can be improved. This paper provides a framework for doing that.

This paper assumes \emph{short packet} transmission. Furthermore, we assume that neither channel gains nor relative phase are known \emph{a priori}, i.e., they need to be estimated by the receiver from the short packet itself. To avoid excessive overhead, we assume such short packets do not have preambles and therefore we cannot use preambles to estimate the channel gains and the relative phase. A challenge is how to estimate the channel gains and the relative phase from the magnitudes of data samples. To address this challenge, we design an overall noncoherent detection system consisting of two components: 1) a channel gains estimator that estimates channel gains without preambles; 2) a detector that builds on top of the estimated channel gains to jointly estimate relative phase and detect data using a belief propagation algorithm. Numerical results show our overall detector performs nearly as well as a ``fictitious'' optimal detector that has perfect knowledge of channel gains and relative phase.

Although this paper focuses on PNC with FSK  modulation, we believe the insight of this paper applies generally to noncoherent detection in other multiuser systems with other modulations. Specifically, our insight is that the relative phase of overlapped signals affects the signal magnitude in multiuser systems, but fortunately the relative phase can be deduced from the magnitudes and this knowledge can be used to improve detection performance.

The remainder of this paper is organized as follows: Section II overviews related work. Section III introduces our system model. We present our design in stages so that our framework can be understood in a systematic manner. Section IV presents optimal detectors with known channel gains. Specifically, Subsections IV-B and IV-C first present an optimal detector for power-balanced channels. Subsection IV-D then follows up with an optimal detector for power-imbalanced channels. Section V gives the ultimate detector design for power-imbalanced channels with unknown channel gains. The detectors in all the two sections estimate the relative phase to improve detection performance under their respective settings. Numerical results on detection performance are given in Subsections of IV-C and V-C. Section VI concludes this paper. The notations of this paper are summarized in Appendix A.

\section{Related Work}
There has been prior investigations on FSK-PNC detection: \cite{yu2016physical} studied coherent detection and \cite{sorensen2009physical}\cite{valenti2011noncoherent} studied noncoherent detection. For coherent detection, the coherent detector has access to not only the magnitude but also the phase of the received signals; for noncoherent detection, the noncoherent detector has only access to the magnitude of the received signals. In this paper, we focus on the noncoherent detection. The investigations on noncoherent detection\cite{sorensen2009physical}\cite{valenti2011noncoherent} did not make use of the fundamental fact that the signal magnitudes in FSK-PNC do contain information about the relative phase and that the relative phase can be deduced from the signal magnitudes to improve detection performance. Our current paper is an attempt to do so. We summarize the work of \cite{sorensen2009physical} and \cite{valenti2011noncoherent} on noncoherent FSK-PNC in the following:

\subsection{Noncoherent Detector in \cite{sorensen2009physical}}
The authors of \cite{sorensen2009physical} put forth a noncoherent detector for power-balanced channels, assuming the detector has perfect knowledge of channel gains. The detector in \cite{sorensen2009physical} detects symbols by marginalizing the relative phase between $\left[ {{\rm{0,2}}\pi } \right]$ in each symbol duration, assuming uniform relative phase distribution (i.e., assuming zero knowledge of the relative phase). By contrast, we will show in our paper here that we can in fact derive the relative phase from the signal magnitudes and that this knowledge can be used to improve the detection performance. In addition, our noncoherent detector can be applied in the more general scenarios with power-imbalanced channels.

\subsection{Noncoherent Detector in \cite{valenti2011noncoherent}}
Ref. \cite{valenti2011noncoherent} gave two noncoherent detectors for two scenarios: 1) detection with channel-gain information and 2) detection without channel-gain information. For 2), by saying ``without the channel-gain information'', we mean the detector in \cite{valenti2011noncoherent} does not know the explicit values of channel gains; the detector in \cite{valenti2011noncoherent} does, however, know the distribution of channel gains in advance.

\textbf{Detector with channel-gain information}: Ref. \cite{valenti2011noncoherent} first designed a channel gain estimator. As in our paper, no preamble is assumed for channel gains estimation. However, in \cite{valenti2011noncoherent}, both phase and magnitude are assumed to be available in each received symbol, while our paper assumes that only the magnitude is available. This channel-gain estimator in\cite{valenti2011noncoherent} estimates the channel gains from the received symbols, and then use the channel gain thus estimated to do noncoherent detection based on signal magnitudes only. In particular, the phase information is exploited in channel-gain estimation, but not in data detection. It is not clear why \cite{valenti2011noncoherent} chose not to use the phase information in data detection: if phase is available, a ``coherent'' detector with better performance could have been designed for data detection. For our paper here, we assume throughout that only signal magnitudes are available from the simple noncoherent-envelope detection circuitry.

\textbf{Detector without channel-gain information}: The second noncoherent detector in \cite{valenti2011noncoherent} does not assume the availability of phase information in the received symbols. As in our paper, only signal magnitudes are available. Unlike our paper, however, this detector in \cite{valenti2011noncoherent} does not try to deduce the relative phase from the signal magnitudes. Instead, it simply approximates the relative phase to be ${\pi  \mathord{\left/{\vphantom {\pi  {\rm{2}}}} \right.\kern-\nulldelimiterspace} {\rm{2}}}$ (i.e., halfway between total constructive interference and total destructive interference when the two users transmit on the same frequency). Based on this approximated phase, it then detects data by marginalizing over channel gains, assuming that the channel-gain distribution and the mean of channel gains power are known in the marginalization process. As will be shown in this paper, the knowledge of the relative phase, which can be deduced from the signal magnitudes, can improve the performance of noncoherent FSK-PNC detection. Another difference of our paper is that we do not assume the knowledge of channel-gain distribution; our ultimate design in Section V simply estimates the channel gains also from the signal magnitudes.

\section{System Model}
 In this paper, for concreteness and as a reference, we assume the bandwidth of our communication system is 1 MHz. Furthermore, we assume the short packets are of 128 bits in size, and thus the packet duration is 128 $\mu s$. In addition, the RFs at users \emph{A} and \emph{B} are not synchronized to a common clock. In general, the phase of the RF at user $u \in \left\{ {A,B} \right\}$ can be expressed as
\begin{align}
\theta _u^{{\rm{RF}}}\left( t \right) = 2\pi \int_{\rm{0}}^t {f_u^{{\rm{RF}}}\left( \tau  \right)d\tau }  + \varphi _u^{{\rm{RF}}} + \varepsilon _u^{{\rm{RF}}}(t) \label{eq:Sys_1}
\end{align}
where $f_u^{{\rm{RF}}}\left( t \right)$ is the RF frequency of user \emph{u}; $\varphi _u^{{\rm{RF}}}$ is an initial phase of the RF (the phase at the beginning of a packet); and $\varepsilon _u^{{\rm{RF}}}(t)$ is a random phase diffusion due to phase noise.

The frequency $f_u^{{\rm{RF}}}\left( t \right)$ may vary from time to time due to the instability and inaccuracy of the frequency-generating oscillator at user \emph{u}. However, for short packets of our interest here, the frequency $f_u^{{\rm{RF}}}\left( t \right)$ remains more or less constant within the packet duration of 128 $\mu s$\cite{wang2018dcap} so that we can write $f_u^{{\rm{RF}}}\left( t \right) = f_u^{{\rm{RF}}}$ for a particular packet. Furthermore, the additional phase due to random phase noise may not have accumulated during the short packet duration so that we can assume $\varepsilon _u^{{\rm{RF}}}(t) = 0$ for a particular packet. In short, we assume that the coherence time of the RF of user \emph{u} is much larger than 128 $\mu s$. Thus, for a particular packet, ~\eqref{eq:Sys_1} can be written as
\begin{align}
\theta _u^{{\rm{RF}}}\left( t \right) = 2\pi f_u^{{\rm{RF}}}t + \varphi _u^{{\rm{RF}}} \label{eq:Sys_2}
\end{align}

Since the  frequency oscillators at  users \emph{A} and \emph{B} operate independently, for the communication between \emph{A} and \emph{B}, there is an initial relative phase $\varphi _B^{{\rm{RF}}} - \varphi _A^{{\rm{RF}}}$ (the relative phase between \emph{A} and \emph{B} at the beginning of a packet) and a \emph{carry frequency offset} (CFO) $f_B^{{\rm{RF}}} - f_A^{{\rm{RF}}}$ between their RFs.

For binary FSK-PNC, user $u \in \left\{ {A,B} \right\}$  employs two frequencies  ${f_{1,u}} = f_u^{{\rm{RF}}} - \Delta f$ and ${f_{2,u}} = f_u^{{\rm{RF}}} + \Delta f$  to transmit bits 0 and 1 respectively, where $2\Delta f = {f_{2,u}} - {f_{1,u}}$   is the frequency separation between ${f_{1,u}}$  and ${f_{2,u}}$. Due to CFO between the RFs of \emph{A} and \emph{B}, the first transmitted frequency ${f_{1,A}}$  at user \emph{A} may not be exactly equal to the first transmitted frequency ${f_{1,B}}$ at user \emph{B}. Likewise for the second frequency. In other words, the CFO is
\begin{align}
{f_{1,B}} - {f_{1,A}} = {f_{2,B}} - {f_{2,A}} = f_B^{{\rm{RF}}} - f_A^{{\rm{RF}}}
\end{align}

Unlike a conventional single-user point-to-point communication system in which there is only one transmitter and one receiver, PNC has two transmitters transmitting simultaneously to one common receiver at the same time. Noncoherent detection in FSK-PNC differs fundamentally from that in conventional point-to-point communication because of the presence of two phenomena in the PNC system:
\begin{enumerate}[(i)]
\item When users \emph{A} and \emph{B} both transmit on the first frequency, the relative phase of their RF carriers, $\theta _B^{{\rm{RF}}}\left( t \right) - \theta _A^{{\rm{RF}}}\left( t \right)$, will have an important impact on the decoding performance because constructive-destructive interference may affect the magnitudes of the signal being received.
\item The CFO $f_B^{{\rm{RF}}} - f_A^{{\rm{RF}}}$ between \emph{A} and \emph{B} will cause the relative phase to vary within a packet, and hence the signal magnitudes also vary within a packet.
\end{enumerate}

We emphasize that what is important to noncoherent detection in PNC is the relative phase and its variation (due to CFO between users \emph{A} and \emph{B}). This will be further elaborated later in this section.
\begin{figure}[ht]
  \centering
  \includegraphics[scale=1.0]{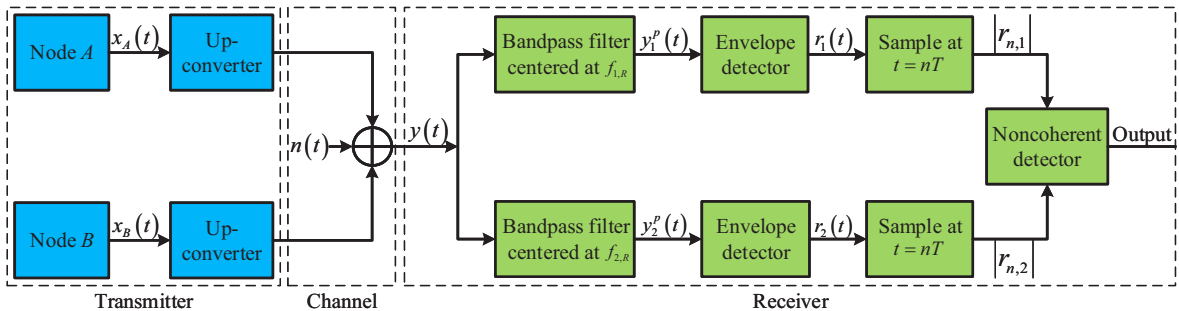}\\
  \caption{Structure of a noncoherent FSK-PNC communication system.}\label{Fig2}
\end{figure}

Fig.\ref{Fig2} shows the structure of a FSK-PNC communication system. The overall structure of the PNC noncoherent receiver is the same as that of a conventional single-user noncoherent receiver (see \cite{CouchLeonW2013Daac}) except for the noncoherent detector at the far right in Fig.\ref{Fig2}. Specifically, as in \cite{CouchLeonW2013Daac}, only the signal magnitudes (i.e., envelopes) of the two frequencies are presented to the noncoherent detector. This paper considers several designs for the noncoherent detector  with progressive generality, the details of which will be presented in Sections IV and V.  In the following, we overview the various processes in Fig.\ref{Fig2}.

\subsection{Baseband Modulator}
This paper assumes both users \emph{A} and \emph{B} adopt continuous-phase FSK\cite{goldsmith2005wireless}. Let ${s_{n,u}} \in \{ 0,1\} $, $n = 0 \cdots N - 1$, be user \emph{u}'s information source symbols, where $N$ is the packet length. Within the symbol duration $nT \le t < (n + 1)T$, the continuous-phase FSK modulated baseband signal of user \emph{u} can be expressed as
\begin{align}
 {x_u}\left( t \right)=
  \begin{cases}
   {e^{ - j2\pi \Delta f(t - nT) + j\varphi _{n,u}^{{\rm{CFSK}}}}}& \text{if ${s_{n,u}} = 0$}\\
   {e^{j2\pi \Delta f(t - nT) + j\varphi _{n,u}^{{\rm{CFSK}}}}}& \text{if ${s_{n,u}} = 1$}
  \end{cases} \label{eq:II-4}
\end{align}
In ~\eqref{eq:II-4}, the baseband signal uses frequency ${\rm{ - }}\Delta f$ to represent bit 0 and frequency $\Delta f$ to represent bit 1. This paper follows the practice of single-user point-to-point noncoherent FSK communication and set $\Delta f = \frac{1}{{2T}}$\cite{goldsmith2005wireless}. In ~\eqref{eq:II-4}, $\varphi _{n,u}^{{\rm{CFSK}}}$ is the phase accumulated over the past \emph{n} symbol periods since continuous phase FSK is applied. As such, $\varphi _{0,u}^{{\rm{CFSK}}} = 0$; $\varphi _{n,u}^{{\rm{CFSK}}} = 2\pi \Delta fT\sum\limits_{i = 0}^{n - 1} {(2{s_{i,u}} - 1)}$ for $n \ge 1$.

\subsection{Upconverter}
The FSK modulated baseband signal ${x_u}\left( t \right)$ for user $u \in \left\{ {A,B} \right\}$ is upconverted to the RF passband. The output of the upconverter within the symbol duration $nT \le t < (n + 1)T$ is
\begin{align}
{\tilde x_u}\left( t \right) = \sqrt {{{\rm{2}} \mathord{\left/{\vphantom {{\rm{2}} T}} \right.
 \kern-\nulldelimiterspace} T}} {\mathop{\Re}\nolimits} \left( {{x_u}\left( t \right){e^{j\left( {2\pi f_u^{{\rm{RF}}}t + \varphi _u^{{\rm{RF}}}} \right)}}} \right) \label{eq:II-5}
\end{align}
where $f_u^{{\rm{RF}}}$ is the RF frequency at user \emph{u}, $\varphi _u^{{\rm{RF}}}$ is the RF's initial phase at user \emph{u}, and the term $\sqrt {{2 \mathord{\left/{\vphantom {2 T}} \right. \kern-\nulldelimiterspace} T}}$ is for normalizing the power. With FSK, when the transmitted bit is 0 at user \emph{u}, the center frequency of the passband signal is ${f_{1,u}} = f_u^{{\rm{RF}}} - \Delta f$; when the transmitted bit is 1 at user \emph{u}, the center frequency of the passband signal is ${f_{2,u}} = f_u^{{\rm{RF}}} + \Delta f$.

\subsection{Bandpass Filter}
In the uplink of PNC, users \emph{A} and \emph{B} transmit their messages to relay \emph{R} simultaneously. Assuming the symbol arrival times are aligned, the received superimposed signal at \emph{R} in the duration $nT \le t < {\rm{(}}n + {\rm{1)}}T$ is
\begin{align}
y\left( t \right) = \sqrt {{{\rm{2}} \mathord{\left/
 {\vphantom {{\rm{2}} T}} \right.\kern-\nulldelimiterspace} T}} \sum\limits_{u \in \left\{ {A,B} \right\}} {\left| {{h_u}} \right|{\mathop{\Re}\nolimits} \left( {{x_u}\left( t \right){e^{j\left( {2\pi f_u^{{\rm{RF}}}t + \varphi _u^{{\rm{RF}}} + {\varphi _{{h_u}}}} \right)}}} \right)}  + n(t) \label{eq:II-6}
\end{align}
where ${\varphi _{{h_u}}}$ is the phase of the channel ${h_u}$, $\left| {{h_u}} \right|$ is the channel gain of channel ${h_u}$, and $n(t)$ is white Gaussian noise with mean zero and double-sided \emph{power spectral density} (PSD) ${{{N_{\rm{0}}}} \mathord{\left/{\vphantom {{{N_{\rm{0}}}} 2}} \right. \kern-\nulldelimiterspace} 2}$. In this paper, we assume the channel to be flat slow-fading: specifically channel ${h_u}$ is constant within one packet duration. In the rest of this paper, for notation simplicity, we write ${\varphi _u} = \varphi _u^{{\rm{RF}}} + {\varphi _{{h_u}}}$.

As shown in Fig.\ref{Fig2}, the received signal $y\left( t \right)$ is then branched off and passed through two bandpass filters with central frequencies ${f_{1,R}}$ in the upper branch and ${f_{2,R}}$ in the lower branch. As in \cite{CouchLeonW2013Daac}, we assume that the bandwidth of the bandpass filter is ${1 \mathord{\left/{\vphantom {1 T}} \right.\kern-\nulldelimiterspace} T}$. Note that, due to small inaccuracies, the center frequencies of the bandpass filters at \emph{R} may not align exactly with the center frequencies of the signals from users \emph{A} and \emph{B}. For example, in the upper branch ${f_{1,R}}$ may not exactly align with the center frequency ${f_{1,A}}$  of user \emph{A} or center frequency  ${f_{1,B}}$ of user \emph{B}. Likewise for the lower branch. To account for this misalignment, the bandwidth of the bandpass filters at \emph{R} may be set to be slightly larger than ${1 \mathord{\left/{\vphantom {1 T}} \right.\kern-\nulldelimiterspace} T}$ (e.g., set to be ${1 \mathord{\left/{\vphantom {1 T}} \right.\kern-\nulldelimiterspace} T} +$maximum misalignment) to allow the whole transmitted signal to pass through.

For our investigation in this paper, however, the misalignment is in the order of 1 kHz to 10 kHz. This is less than two orders of magnitude compared with the symbol rate ${1 \mathord{\left/{\vphantom {1 T}} \right.\kern-\nulldelimiterspace} T}$  of 1 MHz. In this case, in the upper branch, the difference between  ${f_{1,R}}$ and the frequency ${f_{1,A}}$  from user \emph{A} is small; and the difference between ${f_{1,R}}$  and ${f_{1,B}}$  is also small. Likewise for the lower branch. For simplicity of exposition, we therefore assume that the bandpass filter still has bandwidth of ${1 \mathord{\left/{\vphantom {1 T}} \right.\kern-\nulldelimiterspace} T}$.

Let ${y^p}\left( t \right) = \left( {y_1^p\left( t \right),y_2^p\left( t \right)} \right)$ be the outputs of the two bandpass filters, where $y_1^p\left( t \right)$ is the output of the upper bandpass filter in Fig.\ref{Fig2}, and $y_2^p\left( t \right)$ is the output of the lower bandpass filter in Fig.\ref{Fig2}. From ~\eqref{eq:II-6}, ${y^p}\left( t \right)$  in the duration $nT \le t < (n + 1)T$ can be expressed as follows:
\begin{align}
 {y^p}\left( t \right)=
  \begin{cases}
   \left( {\sqrt {\frac{2}{T}} \left| {{h_A}} \right|\cos \left( {{\theta _{1,A}}\left( t \right)} \right) + \sqrt {\frac{2}{T}} \left| {{h_B}} \right|\cos \left( {{\theta _{1,B}}\left( t \right)} \right) + n_1^p(t),n_2^p(t)} \right)& \text{if ${s_{n,A}} = 0,{s_{n,B}} = 0$}\\
   \left( {n_1^p(t),\sqrt {\frac{2}{T}} \left| {{h_A}} \right|\cos \left( {{\theta _{2,A}}\left( t \right)} \right) + \sqrt {\frac{2}{T}} \left| {{h_B}} \right|\cos \left( {{\theta _{2,B}}\left( t \right)} \right) + n_2^p(t)} \right){\kern 1pt}& \text{if ${s_{n,A}} = 1,{s_{n,B}} = 1$}\\
   \left( {\sqrt {\frac{2}{T}} \left| {{h_A}} \right|\cos \left( {{\theta _{1,A}}\left( t \right)} \right) + n_1^p(t),\sqrt {\frac{2}{T}} \left| {{h_B}} \right|\cos \left( {{\theta _{2,B}}\left( t \right)} \right) + n_2^p(t)} \right)& \text{if ${s_{n,A}} = 0,{s_{n,B}} = 1$}\\
   \left( {\sqrt {\frac{2}{T}} \left| {{h_B}} \right|\cos \left( {{\theta _{1,B}}\left( t \right)} \right) + n_1^p(t),\sqrt {\frac{2}{T}} \left| {{h_A}} \right|\cos \left( {{\theta _{2,A}}\left( t \right)} \right) + n_2^p(t)} \right)& \text{if ${s_{n,A}} = 1,{s_{n,B}} = 0$}
  \end{cases} \label{eq:II-7}
\end{align}
where
\begin{align}
{\theta _{1,u}}\left( t \right) =  - 2\pi \Delta f\left( {t - nT} \right) + \varphi _{n,u}^{{\rm{CFSK}}} + 2\pi f_u^{{\rm{RF}}}t + {\varphi _u} = 2\pi {f_{1,u}}t + 2\pi \Delta fnT + \varphi _{n,u}^{{\rm{CFSK}}} + {\varphi _u}
\end{align}
and
\begin{align}
{\theta _{2,u}}\left( t \right) = 2\pi {f_{2,u}}t - 2\pi \Delta fnT + \varphi _{n,u}^{{\rm{CFSK}}} + {\varphi _u}
\end{align}
$n_1^p(t)$ is the bandpass noise in the upper branch with double-sided PSD of $N_0/2$ within the passband of the bandpass filter; $n_2^p(t)$ is the bandpass noise of the lower branch with double-sided PSD of $N_0/2$ within the passband of the lower bandpass filter.

\subsection{Envelope Detector}
After the bandpass filters, two envelope detectors detect the envelopes (i.e., magnitudes) of the passband signals $y_1^p\left( t \right)$ and $y_2^p\left( t \right)$.  In what follows, we express the envelope detection process mathematically. We focus on the case of ${s_{n,A}} = 0$ and ${s_{n,B}} = 0$ to illustrate the basic idea. When ${s_{n,A}} = 0$ and ${s_{n,B}} = 0$, at the output of the upper bandpass filter in Fig.\ref{Fig2}, we have
\begin{align}
y_1^p\left( t \right) = \sqrt {{{\rm{2}} \mathord{\left/{\vphantom {{\rm{2}} T}} \right.
 \kern-\nulldelimiterspace} T}} \left| {{h_A}} \right|\cos \left( {{\theta _{1,A}}\left( t \right)} \right) + \sqrt {{{\rm{2}} \mathord{\left/{\vphantom {{\rm{2}} T}} \right.\kern-\nulldelimiterspace} T}} \left| {{h_B}} \right|\cos \left( {{\theta _{1,B}}\left( t \right)} \right) + n_1^p(t) \label{eq:II-8}
\end{align}
where
\begin{align}
{\theta _{1,A}}\left( t \right) = 2\pi {f_{1,A}}t + 2\pi \Delta fnT + \varphi _{n,A}^{{\rm{CFSK}}} + {\varphi _A}
\end{align}
and
\begin{align}
{\theta _{1,B}}\left( t \right) = 2\pi {f_{1,B}}t + 2\pi \Delta fnT + \varphi _{n,B}^{{\rm{CFSK}}} + {\varphi _B}
\end{align}

\begin{figure}[ht]
  \centering
  \subfigure[PSD of $n_1^p\left( t \right)$.]
  {
  \label{Fig3a}
  \includegraphics[width=0.35\columnwidth]{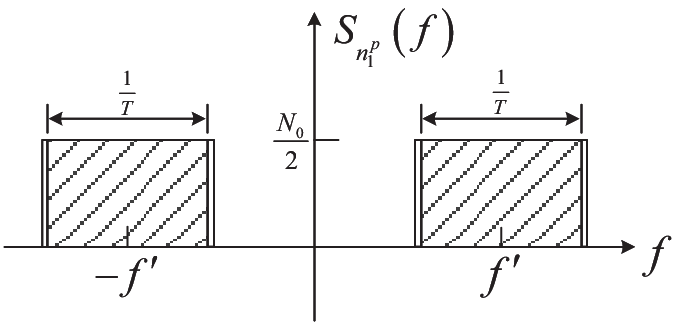}
  }
  \subfigure[PSD of $w_1^i\left( t \right)$.]
  {
  \label{Fig3b}
  \includegraphics[width=0.2\columnwidth]{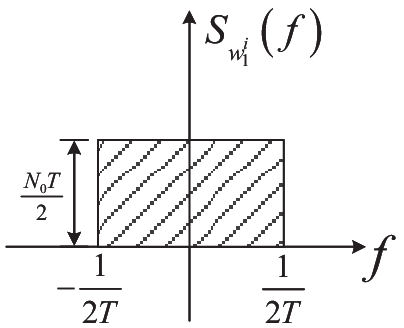}
  }
  \subfigure[PSD of $w_1^q\left( t \right)$.]
  {
  \label{Fig3c}
  \includegraphics[width=0.2\columnwidth]{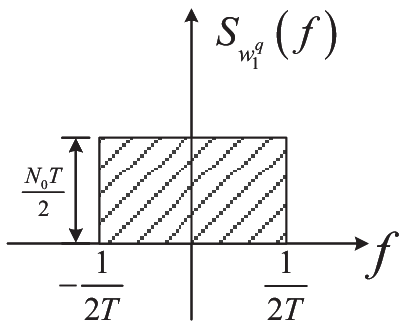}
  }
  \caption{PSDs of  $n_1^p(t)$, $w_1^i\left( t \right)$, and $w_1^q\left( t \right)$. In (a), the bandwidth of the bandpass filter is set to be slightly larger than  $1/T$  to account for potential misalignment of the center frequencies of users \emph{A}, \emph{B}, and \emph{R}. As explained in Subsection III-C, this extra needed bandwidth is minuscule for the inaccuracy typically seen in practical oscillators. For simplicity of exposition, we therefore assume here that the bandwidth of the bandpass filter is still $1/T$, i.e., the shaded part in figure (a).}
\end{figure}

Let
\begin{align}
f' = {{\left( {{f_{1,B}}{\rm{ + }}{f_{1,A}}} \right)} \mathord{\left/
 {\vphantom {{\left( {{f_{1,B}}{\rm{ + }}{f_{1,A}}} \right)} 2}} \right.
 \kern-\nulldelimiterspace} 2}
\end{align}
i.e., $f'$ is the ``average'' of the center frequencies of the signals from \emph{A} and \emph{B}. Also let
\begin{align}
f'' = {{\left( {{f_{1,B}} - {f_{1,A}}} \right)} \mathord{\left/
 {\vphantom {{\left( {{f_{1,B}} - {f_{1,A}}} \right)} 2}} \right.
 \kern-\nulldelimiterspace} 2}
\end{align}
\begin{align}
{\varphi '_n} = {{\left( {2n\pi \Delta fT + \varphi _{n,B}^{{\rm{CFSK}}} + {\varphi _B} + 2n\pi \Delta fT + \varphi _{n,A}^{{\rm{CFSK}}} + {\varphi _A}} \right)} \mathord{\left/
 {\vphantom {{\left( {2n\pi \Delta fT + \varphi _{n,B}^{{\rm{CFSK}}} + {\varphi _B} + 2n\pi \Delta fT + \varphi _{n,A}^{{\rm{CFSK}}} + {\varphi _A}} \right)} 2}} \right.
 \kern-\nulldelimiterspace} 2}
\end{align}
and
\begin{align}
{\varphi ''_n} = {{\left( {\varphi _{n,B}^{{\rm{CFSK}}} + {\varphi _B} - \varphi _{n,A}^{{\rm{CFSK}}} - {\varphi _A}} \right)} \mathord{\left/
 {\vphantom {{\left( {\varphi _{n,B}^{{\rm{CFSK}}} + {\varphi _B} - \varphi _{n,A}^{{\rm{CFSK}}} - {\varphi _A}} \right)} 2}} \right.
 \kern-\nulldelimiterspace} 2}
\end{align}
With the above notations, ${\theta _{1,A}}\left( t \right)$ and ${\theta _{1,B}}\left( t \right)$ in ~\eqref{eq:II-8} can be rewritten as
\begin{align}
{\theta _{1,A}}\left( t \right) = 2\pi \left( {f' - f''} \right)t + {\varphi '_n} - {\varphi ''_n}
\end{align}
and
\begin{align}
{\theta _{1,B}}\left( t \right) = 2\pi \left( {f' + f''} \right)t + {\varphi '_n} + {\varphi ''_n}
\end{align}
Then $y_1^p\left( t \right)$ in ~\eqref{eq:II-8} can be further expanded as
\begin{equation}
\begin{split}
&y_1^p\left( t \right)\\
&= \sqrt {{2 \mathord{\left/{\vphantom {2 T}} \right.\kern-\nulldelimiterspace} T}} \left[ {\left| {{h_A}} \right|\cos \left( {2\pi \left( {f' - f''} \right)t + {{\varphi '}_n} - {{\varphi ''}_n}} \right) + \left| {{h_B}} \right|\cos \left( {2\pi \left( {f' + f''} \right)t + {{\varphi '}_n} + {{\varphi ''}_n}} \right)} \right] + n_1^p(t)\\
&= \sqrt {{2 \mathord{\left/{\vphantom {2 T}} \right.\kern-\nulldelimiterspace} T}} \left| {{h_A}} \right|\left[ {\cos \left( {2\pi f''t + {{\varphi ''}_n}} \right)\cos \left( {2\pi f't + {{\varphi '}_n}} \right) + \sin \left( {2\pi f''t + {{\varphi ''}_n}} \right)\sin \left( {2\pi f't + {{\varphi '}_n}} \right)} \right] \\
&+ \sqrt {{2 \mathord{\left/{\vphantom {2 T}} \right.\kern-\nulldelimiterspace} T}} \left| {{h_B}} \right|\left[ {\cos \left( {2\pi f''t + {{\varphi ''}_n}} \right)\cos \left( {2\pi f't + {{\varphi '}_n}} \right) - \sin \left( {2\pi f''t + {{\varphi ''}_n}} \right)\sin \left( {2\pi f't + {{\varphi '}_n}} \right)} \right] + n_1^p(t)\\
&= \sqrt {{2 \mathord{\left/{\vphantom {2 T}} \right.\kern-\nulldelimiterspace} T}} \left( {\left| {{h_A}} \right| + \left| {{h_B}} \right|} \right)\cos \left( {{{\tilde \theta \left( t \right)} \mathord{\left/{\vphantom {{\tilde \theta \left( t \right)} 2}} \right.
 \kern-\nulldelimiterspace} 2}} \right)\cos \left( {2\pi f't + {{\varphi '}_n}} \right) \\
&+ \sqrt {{2 \mathord{\left/{\vphantom {2 T}} \right.\kern-\nulldelimiterspace} T}} \left( {\left| {{h_A}} \right| - \left| {{h_B}} \right|} \right)\sin \left( {{{\tilde \theta \left( t \right)} \mathord{\left/{\vphantom {{\tilde \theta \left( t \right)} 2}} \right.\kern-\nulldelimiterspace} 2}} \right)\sin \left( {2\pi f't + {{\varphi '}_n}} \right) + n_1^p(t)
\end{split} \label{eq:II-9}
\end{equation}
where
\begin{align}
\tilde \theta \left( t \right) = {\theta _{1,B}}\left( t \right) - {\theta _{1,A}}\left( t \right) = 4\pi f''t + 2{\varphi ''_n} \label{eq:II-9.5}
\end{align}
is the relative phase between users \emph{A} and \emph{B}; and  $n_1^p(t)$ is the passband noise. The PSD of  $n_1^p(t)$ is illustrated in Fig.\ref{Fig3a}.

The passband noise $n_1^p(t)$ can be modeled as
\begin{align}
n_1^p(t) = \sqrt {{{\rm{2}} \mathord{\left/{\vphantom {{\rm{2}} T}} \right.
 \kern-\nulldelimiterspace} T}} w_1^i\left( t \right)\cos \left( {2\pi f't + {{\varphi '}_n}} \right) - \sqrt {{{\rm{2}} \mathord{\left/{\vphantom {{\rm{2}} T}} \right.\kern-\nulldelimiterspace} T}} w_1^q\left( t \right)\sin \left( {2\pi f't + {{\varphi '}_n}} \right) \label{eq:II-10}
\end{align}
where $w_1^i\left( t \right)$ and $w_1^q\left( t \right)$ are two independent \emph{wide-sense stationary} (WSS) processes and their PSDs are shown in Fig.\ref{Fig3b} and Fig.\ref{Fig3c}. From ~\eqref{eq:II-10}, the PSD of $n_1^p(t)$ is
\begin{align}
{S_{n_1^p}}\left( f \right) = \frac{1}{{2T}}{S_{w_1^i}}\left( {f - f'} \right) + \frac{1}{{2T}}{S_{w_1^i}}\left( {f + f'} \right) + \frac{1}{{2T}}{S_{w_1^q}}\left( {f - f'} \right) + \frac{1}{{2T}}{S_{w_1^q}}\left( {f + f'} \right) \label{eq:II-11}
\end{align}
where ${S_{w_1^i}}\left( f \right)$ and ${S_{w_1^q}}\left( f \right)$ are PSDs of $w_1^i\left( t \right)$ and $w_1^q\left( t \right)$ respectively. From ~\eqref{eq:II-11} and the PSDs ${S_{w_1^i}}\left( f \right)$ and ${S_{w_1^q}}\left( f \right)$ in Fig.\ref{Fig3b} and Fig.\ref{Fig3c}, we can retrieve the exact PSD of $n_1^p(t)$ in Fig.\ref{Fig3a}. This validates our model of $n_1^p(t)$ in ~\eqref{eq:II-10}.

For a specific point in time ${t_0}$, since the expectation of $n_1^p({t_0})$ is $E\left( {n_1^p({t_0})} \right) = 0$, we have $E\left( {w_1^i\left( {{t_0}} \right)} \right) = E\left( {w_1^q\left( {{t_0}} \right)} \right) = 0$. Furthermore, the variance of $w_1^i\left( {{t_0}} \right)$ is
\begin{align}
{\rm{Var}}\left( {w_1^i\left( {{t_0}} \right)} \right) = \int_{ - \infty }^{ + \infty } {{S_{w_1^i}}\left( f \right)df}  = {{{N_0}} \mathord{\left/{\vphantom {{{N_0}} 2}} \right.\kern-\nulldelimiterspace} 2}
\end{align}
Similarly, ${\rm{Var}}\left( {w_1^q\left( {{t_0}} \right)} \right) = {{{N_0}} \mathord{\left/{\vphantom {{{N_0}} 2}} \right.\kern-\nulldelimiterspace} 2}$.

Substituting ~\eqref{eq:II-10} into ~\eqref{eq:II-9}, we have
\begin{equation}
\begin{split}
y_1^p\left( t \right) &= \sqrt {{{\rm{2}} \mathord{\left/
 {\vphantom {{\rm{2}} T}} \right.
 \kern-\nulldelimiterspace} T}} \left[ {\left( {\left| {{h_A}} \right| + \left| {{h_B}} \right|} \right)\cos \left( {{{\tilde \theta \left( t \right)} \mathord{\left/
 {\vphantom {{\tilde \theta \left( t \right)} 2}} \right.
 \kern-\nulldelimiterspace} 2}} \right) + w_1^i\left( t \right)} \right]\cos \left( {2\pi f't + {{\varphi '}_n}} \right) \\
 &+ \sqrt {{{\rm{2}} \mathord{\left/
 {\vphantom {{\rm{2}} T}} \right.
 \kern-\nulldelimiterspace} T}} \left[ {\left( {\left| {{h_A}} \right| - \left| {{h_B}} \right|} \right)\sin \left( {{{\tilde \theta \left( t \right)} \mathord{\left/
 {\vphantom {{\tilde \theta \left( t \right)} 2}} \right.
 \kern-\nulldelimiterspace} 2}} \right) - w_1^q\left( t \right)} \right]\sin \left( {2\pi f't + {{\varphi '}_n}} \right)
\end{split}
\end{equation}
The envelope of $y_1^p\left( t \right)$ is the square root of the sum of the squared coefficients of $\cos \left( {2\pi f't + {{\varphi '}_n}} \right)$ and $\sin \left( {2\pi f't + {{\varphi '}_n}} \right)$. At the envelope detector, the envelope of $y_1^p\left( t \right)$ is normalized by $\sqrt{T/2}$. Let ${r_1}\left( t \right)$ be the output of the envelope detector at the upper branch. In the duration $nT \le t < (n + 1)T$,
\begin{equation}
\begin{split}
{r_1}\left( t \right) &= \sqrt {{{\left[ {\left( {\left| {{h_A}} \right| + \left| {{h_B}} \right|} \right)\cos \left( {{{\tilde \theta \left( t \right)} \mathord{\left/
 {\vphantom {{\tilde \theta \left( t \right)} 2}} \right.
 \kern-\nulldelimiterspace} 2}} \right) + w_1^i\left( t \right)} \right]}^2} + {{\left[ {\left( {\left| {{h_A}} \right| - \left| {{h_B}} \right|} \right)\sin \left( {{{\tilde \theta \left( t \right)} \mathord{\left/
 {\vphantom {{\tilde \theta \left( t \right)} 2}} \right.
 \kern-\nulldelimiterspace} 2}} \right) - w_1^q\left( t \right)} \right]}^2}} \\
 &= \left| {\left| {{h_A}} \right|{e^{ - j{{\tilde \theta \left( t \right)} \mathord{\left/
 {\vphantom {{\tilde \theta \left( t \right)} 2}} \right.
 \kern-\nulldelimiterspace} 2}}} + \left| {{h_B}} \right|{e^{j{{\tilde \theta \left( t \right)} \mathord{\left/
 {\vphantom {{\tilde \theta \left( t \right)} 2}} \right.
 \kern-\nulldelimiterspace} 2}}} + {w_1}\left( t \right)} \right|
\end{split}
\end{equation}
where we write $w_1^{}\left( t \right) = w_1^i\left( t \right) + jw_1^q\left( t \right)$.

Then, ${r_1}\left( t \right)$ is sampled at $t = nT$, $n = 0, \cdots, N - 1$. Let $\left| {{r_{n,1}}} \right| = {r_1}\left( {nT} \right)$, we have
\begin{align}
\left| {{r_{n,1}}} \right| = {r_1}\left( {nT} \right) = \left| {\left| {{h_A}} \right|{e^{ - j{{{{\tilde \theta }_n}} \mathord{\left/
 {\vphantom {{{{\tilde \theta }_n}} 2}} \right.
 \kern-\nulldelimiterspace} 2}}} + \left| {{h_B}} \right|{e^{j{{{{\tilde \theta }_n}} \mathord{\left/
 {\vphantom {{{{\tilde \theta }_n}} 2}} \right.
 \kern-\nulldelimiterspace} 2}}} + {w_{n,1}}} \right| \label{eq:II-12}
\end{align}
where ${w_{n,1}} = {w_1}\left( {nT} \right) \sim CN(0,{N_0})$, and ${\tilde \theta _n} = \tilde \theta \left( {nT} \right)$ is the relative phase between users \emph{A} and \emph{B} at $t = nT$.

Continuing from ~\eqref{eq:II-9.5}, let
\begin{align}
\tilde f = 2f'' = {f_{1,B}} - {f_{1,A}} = f_B^{{\rm{RF}}} - \Delta f - f_A^{{\rm{RF}}} + \Delta f = f_B^{{\rm{RF}}} - f_A^{{\rm{RF}}}
\end{align}
be CFO between \emph{A} and \emph{B}, and $\tilde \varphi  = {\varphi _B} - {\varphi _A}$ be the initial relative phase between \emph{A} and \emph{B}. At $t = nT$, the relative phase can be expressed as
\begin{align}
{\tilde \theta _n} = \tilde \theta \left( {nT} \right) = 2n\pi \tilde fT + \tilde \varphi  + \varphi _{n,B}^{{\rm{CFSK}}} - \varphi _{n,A}^{{\rm{CFSK}}} \label{eq:II-13}
\end{align}
Appendix B shows that the relative phase ${\tilde \theta _n}$ can be further expressed as
\begin{align}
{\tilde \theta _n} = 2n\pi \tilde fT + \tilde \varphi
\end{align}

Similarly, at $t = nT$, the output of the envelope detector at the lower branch is $\left| {{r_{n,2}}} \right| = \left| {{w_{n,2}}} \right|$, where ${w_{n,2}} \sim CN(0,{N_0})$. Let ${{\bf{r}}_n} = \left( {\left| {{r_{n,1}}} \right|,\left| {{r_{n,2}}} \right|} \right)$. In general, we have
\begin{align}
 {{\bf{r}}_n}=
  \begin{cases}
   \left( {\left| {\left| {{h_A}} \right|{e^{ - j{{{{\tilde \theta }_n}} \mathord{\left/
   {\vphantom {{{{\tilde \theta }_n}} 2}} \right.
   \kern-\nulldelimiterspace} 2}}} + \left| {{h_B}} \right|{e^{j{{{{\tilde \theta }_n}} \mathord{\left/
   {\vphantom {{{{\tilde \theta }_n}} 2}} \right.
   \kern-\nulldelimiterspace} 2}}} + {w_{n,1}}} \right|,\left| {{w_{n,2}}} \right|} \right)& \text{if ${s_{n,A}} = 0,{s_{n,B}} = 0$}\\
   \left( {{\kern 1pt} \left| {{w_{n,1}}} \right|,\left| {\left| {{h_A}} \right|{e^{ - j{{{{\tilde \theta }_n}} \mathord{\left/
   {\vphantom {{{{\tilde \theta }_n}} 2}} \right.
   \kern-\nulldelimiterspace} 2}}} + \left| {{h_B}} \right|{e^{j{{{{\tilde \theta }_n}} \mathord{\left/
   {\vphantom {{{{\tilde \theta }_n}} 2}} \right.
   \kern-\nulldelimiterspace} 2}}} + {w_{n,2}}} \right|} \right)& \text{if ${s_{n,A}} = 1,{s_{n,B}} = 1$}\\
   \left( {\left| {\left| {{h_A}} \right|{e^{ - j{{{{\tilde \theta }_n}} \mathord{\left/
   {\vphantom {{{{\tilde \theta }_n}} 2}} \right.
   \kern-\nulldelimiterspace} 2}}} + {w_{n,1}}} \right|,\left| {\left| {{h_B}} \right|{e^{j{{{{\tilde \theta }_n}} \mathord{\left/
   {\vphantom {{{{\tilde \theta }_n}} 2}} \right.
   \kern-\nulldelimiterspace} 2}}} + {w_{n,2}}} \right|} \right)& \text{if ${s_{n,A}} = 0,{s_{n,B}} = 1$}\\
   \left( {{\kern 1pt} \left| {\left| {{h_B}} \right|{e^{j{{{{\tilde \theta }_n}} \mathord{\left/
   {\vphantom {{{{\tilde \theta }_n}} 2}} \right.
   \kern-\nulldelimiterspace} 2}}} + {w_{n,1}}} \right|,\left| {\left| {{h_A}} \right|{e^{ - j{{{{\tilde \theta }_n}} \mathord{\left/
   {\vphantom {{{{\tilde \theta }_n}} 2}} \right.
   \kern-\nulldelimiterspace} 2}}} + {w_{n,2}}} \right|} \right)& \text{if ${s_{n,A}} = 1,{s_{n,B}} = 0$}
  \end{cases} \label{eq:II-14}
\end{align}

\subsection{Noncoherent Detector}
Our noncoherent detector makes decisions based on the magnitudes of the received signals. The optimal decision rule depends on the relative phase ${\tilde \theta _n}$ between users \emph{A} and \emph{B}. Specifically, if two users transmit on the same frequency, the relative phase between user \emph{A} and \emph{B} will cause constrictive-destructive interference on the magnitude of the signal on that frequency. For example, if both users transmit on the first frequency, the received magnitude on the first frequency is
\begin{align}
\left| {{r_{n,1}}} \right| = \sqrt{\begin{array}{l}
{\left| {{h_A}} \right|^{\rm{2}}}{\rm{ + }}{\left| {{h_B}} \right|^{\rm{2}}}{\rm{ + 2}}\left| {{h_A}} \right|\left| {{h_B}} \right|\cos \left( {{{\tilde \theta }_n}} \right) + {\left| {{w_{n,1}}} \right|^2}\\
 + 2\left| {{h_A}} \right|{\mathop{\Re}\nolimits} \left( {{e^{j{{{{\tilde \theta }_n}} \mathord{\left/
 {\vphantom {{{{\tilde \theta }_n}} 2}} \right.
 \kern-\nulldelimiterspace} 2}}}{w_{n,1}}} \right) + 2\left| {{h_B}} \right|{\mathop{\Re}\nolimits} \left( {{e^{ - j{{{{\tilde \theta }_n}} \mathord{\left/
 {\vphantom {{{{\tilde \theta }_n}} 2}} \right.
 \kern-\nulldelimiterspace} 2}}}{w_{n,1}}} \right)
\end{array}}  \label{eq:II-15}
\end{align}
The noise term $2\left| {{h_A}} \right|{\mathop{\Re}\nolimits} \left( {{e^{j{{{{\tilde \theta }_n}} \mathord{\left/{\vphantom {{{{\tilde \theta }_n}} 2}} \right.\kern-\nulldelimiterspace} 2}}}{w_{n,1}}} \right)$ is identically distributed for different relative phase ${\tilde \theta _n}$. Likewise for $2\left| {{h_B}} \right|{\mathop{\Re}\nolimits} \left( {{e^{ - j{{{{\tilde \theta }_n}} \mathord{\left/{\vphantom {{{{\tilde \theta }_n}} 2}} \right.\kern-\nulldelimiterspace} 2}}}{w_{n,1}}} \right)$. However, ${\rm{2}}\left| {{h_A}} \right|\left| {{h_B}} \right|\cos \left( {{{\tilde \theta }_n}} \right)$ is a fixed term depending on the relative phase ${\tilde \theta _n}$, which affects the magnitude $\left| {{r_{n,1}}} \right|$ significantly even in the absence of noise. Thus, the performance of noncoherent detection depends much on the relative phase ${\tilde \theta _n}$.

For a single-user point-to-point noncoherent system consisting of a transmitter \emph{A} and a receiver \emph{R} (i.e., without transmitter \emph{B}), there is no overlapped signals from multiple transmitters, and the phase of the received signal does not affect magnitude and the performance. The design of the noncoherent detector in PNC cannot blindly follow the design of the single-user noncoherent detector. For optimal performance, a PNC noncoherent detector needs to take the relative phase ${\tilde \theta _n}$ between the signals of \emph{A} and \emph{B} into account. Given that we assume short packets without preamble, a challenge is how to extract the knowledge of ${\tilde \theta _n}$ without preamble. We will show that we can estimate ${\tilde \theta _n}$ well by examining only the magnitudes of the received data samples themselves.

\section{Optimal Detector Design with Known Channel Gains}
To bring out the essence of our detectors, we first discuss a set-up in which the channels are power-balanced (i.e., $\left| {{h_A}} \right| = \left| {{h_B}} \right|{\rm{ = 1}}$) and the noncoherent detector has \emph{a priori} knowledge of the balanced channel gains. However, the channel phases ${\varphi _{{h_A}}}$ and ${\varphi _{{h_B}}}$  are not known. We show how the relative phase
\begin{align}
{\tilde \theta _n}=2n\pi \tilde fT + \tilde \varphi=2n\pi \tilde fT + \varphi _B^{{\rm{RF}}} - \varphi _A^{{\rm{RF}}} + {\varphi _{{h_B}}} - {\varphi _{{h_A}}}
\end{align}
can be estimated and be used to improve the performance of the noncoherent detector in FSK-PNC. Subsection IV-D will then extend the discussion to power-imbalanced channels with known channel gains.

The XORed message of \emph{A} and \emph{B} is ${\{ {s_n} = {s_{n,A}} \oplus {s_{n,B}}\} _{n = 0,...,N - 1}}$. Each user independently transmits bits 0 and 1 with equal probabilities so that $\Pr\left( {{s_n} = 0} \right) = \Pr\left( {{s_n} = 1} \right) = {1 \mathord{\left/{\vphantom {1 2}} \right.\kern-\nulldelimiterspace} 2}$. Let $\Pr\left( {\left. {{s_n}} \right|{\bf{R}}} \right)$ be the conditional \emph{probability density function} (PDF) of the \emph{n}-th XORed symbol given the magnitudes of the received signals of the overall received packet, ${\bf{R}} = \left( {{{\bf{r}}_0}, \cdots, {{\bf{r}}_{N - 1}}} \right)$, where ${{\bf{r}}_n} = \left( {\left| {{r_{n,1}}} \right|,\left| {{r_{n,2}}} \right|} \right)$ . The detector detects the XORed symbol based on the \emph{maximum a posteriori probability} (MAP) criterion:
\begin{align}
s_n^ * {\rm{ = }}\mathop {\arg\max}\limits_{{s_n} = 0,1}\Pr\left( {\left. {{s_n}} \right|{\bf{R}}} \right)  \label{eq:III-1}
\end{align}
where $s_n^ * $ denotes the decision on the XORed symbol ${s_n}$ of \emph{A} and \emph{B}, $n = {\rm{0}}, \cdots ,N - 1$.

As shown in Section III, the relative phase ${\tilde \theta _n}$ between users \emph{A} and \emph{B} induces constructive-destructive interference on the magnitudes of the received signals. To take into account the effect of ${\tilde \theta _n}$, we write
\begin{align}
\Pr\left( {\left. {{s_n}} \right|{\bf{R}}} \right) = \int {\Pr\left( {\left. {{s_n},{{\tilde \theta }_n}} \right|{\bf{R}}} \right)d{{\tilde \theta }_n}}   \label{eq:III-2}
\end{align}
There are two uncertainties in $\Pr\left( {\left. {{s_n},{{\tilde \theta }_n}} \right|{\bf{R}}} \right)$: the XORed symbol ${s_n}$  and the relative phase ${\tilde \theta _n}$. It is the second uncertainty ${\tilde \theta _n}$ that makes noncoherent detection in PNC different from noncoherent detection in a conventional single-user point-to-point system. Unlike a multiuser system such as PNC, the relative phase between two transmitters does not exist in a single-user point-to-point system; although the signal from the single transmitter may still have a phase, there is no constructive-destructive interference and the phase does not affect the signal magnitude.

\subsection{Brief Review of Prior Scheme}
Ref. \cite{sorensen2009physical} also studied the case of power-balanced channel gains with known channel gains. Let us briefly review \cite{sorensen2009physical} to bring out the difference of our approach. Ref. \cite{sorensen2009physical} argued that the noncoherent detector can deal with the dependence of signal magnitude on relative phase ${\tilde \theta _n}$ by marginalizing ${\tilde \theta _n}$ between $[0,2\pi )$ in each symbol period, assuming ${\tilde \theta _n}$ is uniformly distributed over the interval $[0,2\pi )$. Specifically, for all \emph{n}, \eqref{eq:III-2} can be expressed as
\begin{equation}
\begin{split}
\Pr\left( {\left. {{s_n}} \right|{\bf{R}}} \right)&=\int {\Pr\left( {\left. {{s_n},{{\tilde \theta }_n}} \right|{\bf{R}}} \right)d{{\tilde \theta }_n}} \\
&= \int {\Pr\left( {\left. {{s_n},{{\tilde \theta }_n}} \right|{{\bf{r}}_n}} \right)d{{\tilde \theta }_n}}\\
&=\frac{{\Pr\left( {{s_n}} \right)}}{{2\pi \Pr\left( {{{\bf{r}}_n}} \right)}}\int {\Pr\left( {\left. {{{\bf{r}}_n}} \right|{s_n},{{\tilde \theta }_n}} \right)d{{\tilde \theta }_n}}
\end{split} \label{eq:III-A1}
\end{equation}
where $\Pr\left( {\left. {{{\bf{r}}_n}} \right|{s_n}{\rm{ = 0}},{{\tilde \theta }_n}} \right)$ is the conditional PDF of the magnitudes ${{\bf{r}}_n}$ when two users transmit on the same frequency, and $\Pr\left( {\left. {{{\bf{r}}_n}} \right|{s_n}{\rm{ = 1}},{{\tilde \theta }_n}} \right)$  is the conditional PDF of the magnitudes ${{\bf{r}}_n}$ when two users transmit on different frequencies. The first line to the second line in \eqref{eq:III-A1} is valid only if the information of ${s_n}$ and ${\tilde \theta _n}$ is only contained in ${{\bf{r}}_n}$ but not in ${{\bf{r}}_m} \in {\bf{R}}$, $m \ne n$. In \eqref{eq:III-A1}, an implicit assumption is that ${\tilde \theta _n}$ for different \emph{n} are i.i.d. and uniformly distributed. The detector is only optimal if ${\tilde \theta _n}$ changes very quickly from symbol to symbol (i.e., either the channels or the local oscillators of users \emph{A} and \emph{B} have extremely short coherence times), although this assumption was not mentioned explicitly in \cite{sorensen2009physical}. However, this is not the case in most practical systems. For example, for our assumed system with 1 MHz bandwidth, one symbol lasts for 1 $\mu s$ only.  The coherence time of practical channels and the coherence time of local oscillators are typically much larger than that\cite{wang2018dcap}. In this case, the first line to the second line in \eqref{eq:III-A1} is not valid and symbol-by-symbol independent detection in \eqref{eq:III-A1} is not optimal because ${\tilde \theta _n}$ for successive \emph{n} are highly correlated and can be nearly equal for short packets. As will be shown, by observing the signal magnitudes of successive symbols, we can estimate the relative phase. A better detector than that based on \eqref{eq:III-A1} can then be designed. In this paper, we refer to the method in \cite{sorensen2009physical} as \textbf{marginalized-phased detector (MPD)}.

\subsection{Optimal Detector}
Our detector makes decisions taking into account the relationship between the relative phase and signal magnitudes. We assume that the CFO is constant within the short packet. Thus, the relative phase changes from symbol to symbol with a constant increment. To incorporate this relationship, we write \eqref{eq:III-2} as
\begin{align}
\Pr\left( {\left. {{s_n}} \right|{\bf{R}}} \right) = \int {\Pr\left( {\left. {{s_n},{{\tilde \theta }_n},\vartheta } \right|{\bf{R}}} \right)d{{\tilde \theta }_n}} d\vartheta   \label{eq:III-B0}
\end{align}
where $\vartheta  = {\tilde \theta _n} - {\tilde \theta _{n - 1}} = 2\pi \tilde fT$ for $n \in \left[ {1,N - 1} \right]$ is the symbol-to-symbol drift of relative phase induced by a constant CFO $\tilde f$  between users \emph{A} and \emph{B}.

\begin{figure}[ht]
  \centering
  \includegraphics[scale=0.80]{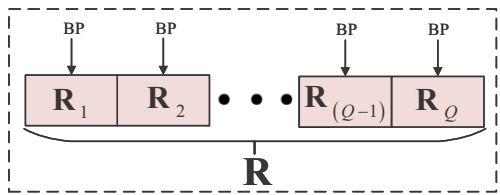}\\
  \caption{The overall received-signal-magnitudes \textbf{R} are divided into \emph{Q} blocks. BP is run over each block ${\textbf{R}_q}$, $q \in \left[ {1,Q} \right]$, rather than the whole \textbf{R}. }\label{Fig3_5}
\end{figure}

The detector is global-optimal because it detects ${s_n}$, ${\tilde \theta _n}$, and $\vartheta $ jointly, not separately, and applies the MAP criterion over the received-signal-magnitudes over the whole received packet \textbf{R}, not just the received-signal-magnitude over single ${{\bf{r}}_n}$. A Belief Propagation (BP) algorithm can be constructed for the computation of $\Pr\left( {\left. {{s_n},{{\tilde \theta }_n},\vartheta } \right|{\bf{R}}} \right)$ \cite{PearlJudea1988Prii}\cite{lu2012asynchronous}\cite{shao2017asynchronous}. The BP algorithm is based on systematic application of Bayes' rule. Although strictly speaking, we need to examine the whole sequence of \textbf{R} for optimal detection of a symbol ${s_n}$, our numerical results in Subsection IV-C suggest that for optimal performance, the detector needs only examine a few successive received signal magnitudes to detect each symbol with high accuracy. That is, we run BP over only a few successive received signal magnitudes adjacent to ${{\bf{r}}_n}$. In particular, we can divide the whole received packet \textbf{R} into \emph{Q} blocks, with each block having \emph{L} received-signal-magnitudes, as shown in Fig.\ref{Fig3_5}. Thus, the packet length is $N = QL$\footnote{When \emph{N} cannot be divided by \emph{Q} in the last block (i.e., the \emph{Q}-th block), we can detect ${s_n}$, ${\tilde \theta _n}$, and $\vartheta $ by borrowing symbols from the previous block, i.e., (\emph{Q}-1)-th block.}. In the \emph{Q}-th block, $q \in \left[ {1,Q} \right]$, \eqref{eq:III-B0} can be rewritten as
\begin{align}
\Pr\left( {\left. {{s_n}} \right|{{\bf{R}}_q}} \right) = \int {\Pr\left( {\left. {{s_n},{{\tilde \theta }_n},\vartheta } \right|{{\bf{R}}_q}} \right)d{{\tilde \theta }_n}} d\vartheta  \label{eq:III-B00}
\end{align}

The integrand $\Pr\left( {\left. {{s_n},{{\tilde \theta }_n},\vartheta } \right|{{\bf{R}}_q}} \right)$ in \eqref{eq:III-B00} can be further expanded. Appendix C derives an expression for general channel gains $\left| {{h_A}} \right|$ and $\left| {{h_B}} \right|$ where $\left| {{h_A}} \right|$ and $\left| {{h_B}} \right|$ are not necessarily equal. We draw on the results from Appendix C in this special case where $\left| {{h_A}} \right| = \left| {{h_B}} \right|{\rm{ = 1}}$. From Appendix C, the integrand in \eqref{eq:III-B00} can be expressed as
\begin{equation}
\begin{split}
&\Pr\left( {\left. {{s_n},{{\tilde \theta }_n},\vartheta } \right|{{\bf{R}}_q}} \right)\\
&= {\eta _q}\Pr\left( \vartheta  \right)\Pr\left( {{{\bf{r}}_n}\left| {{s_n},{{\tilde \theta }_n}} \right.} \right)\int {d{{\tilde \theta }_{n - 1}} \cdots d{{\tilde \theta }_{\left( {q - 1} \right)L}}} \prod\limits_{i = (q - 1)L}^{n - 1} {\sum\limits_{ {s_i}} {\Pr\left( {{{\bf{r}}_i}\left| {{s_i},{{\tilde \theta }_i}} \right.} \right)\delta \left( {{{\tilde \theta }_i} - {{\left[ {{{\tilde \theta }_{i + 1}} - \vartheta } \right]}_{{\rm{2}}\pi }}} \right)} }\\
&\times \int {d{{\tilde \theta }_{n + 1}} \cdots d{{\tilde \theta }_{qL - 1}}} \prod\limits_{i = n + 1}^{qL - 1} {\sum\limits_{{s_i}} {\Pr\left( {{{\bf{r}}_i}\left| {{s_i},{{\tilde \theta }_i}} \right.} \right)\delta \left( {{{\tilde \theta }_i} - {{\left[ {{{\tilde \theta }_{i - 1}} + \vartheta } \right]}_{{\rm{2}}\pi }}} \right)} }
\end{split}\label{eq:III-B1}
\end{equation}
where ${{\bf{R}}_q} = \left( {{{\bf{r}}_{\left( {q - 1} \right)L}}, \cdots ,{{\bf{r}}_{qL - 1}}} \right)$; ${\eta _q}$ is a constant in the \emph{q}-th block; $\vartheta  = {\tilde \theta _n} - {\tilde \theta _{n - 1}} = 2\pi \tilde fT$, $n \in \left[ {1,N - 1} \right]$, is the symbol-to-symbol drift of relative phase induced by a constant CFO $\tilde f$ between users \emph{A} and \emph{B}. The range of CFO investigated in this paper is $\tilde f \in \left[ { - 10{\kern 1pt} {\kern 1pt} {\rm{kHz}},10{\kern 1pt} {\kern 1pt} {\rm{kHz}}} \right]$ (the oscillators in software-defined radio boards, for example, typically have CFO smaller than this range), and hence $\vartheta  \in \left[ { - {\rm{0}}{\rm{.02}}\pi ,{\rm{0}}{\rm{.02}}\pi } \right]$. In \eqref{eq:III-B1}, $\Pr\left( \vartheta  \right)$ is the distribution of $\vartheta $. As a conservative measure, we assume we do not have further information about  $\vartheta $ except that it falls within the said range. In particular, we assume  $\vartheta $ is uniformly distributed within $\left[ { - {\rm{0}}{\rm{.02}}\pi ,{\rm{0}}{\rm{.02}}\pi } \right]$, and outside $\left[ { - {\rm{0}}{\rm{.02}}\pi ,{\rm{0}}{\rm{.02}}\pi } \right]$, $\Pr\left( \vartheta  \right){\rm{ = 0}}$. In \eqref{eq:III-B1}, $\delta \left(  \bullet  \right)$ is the Dirac delta function, and ${\left[  \bullet  \right]_{2\pi }}{\rm{ = }} \bullet \bmod 2\pi$. Note that the range of ${\tilde \theta _n}$   is $[0,2\pi )$.
\begin{figure}[ht]
  \centering
  \includegraphics[scale=0.80]{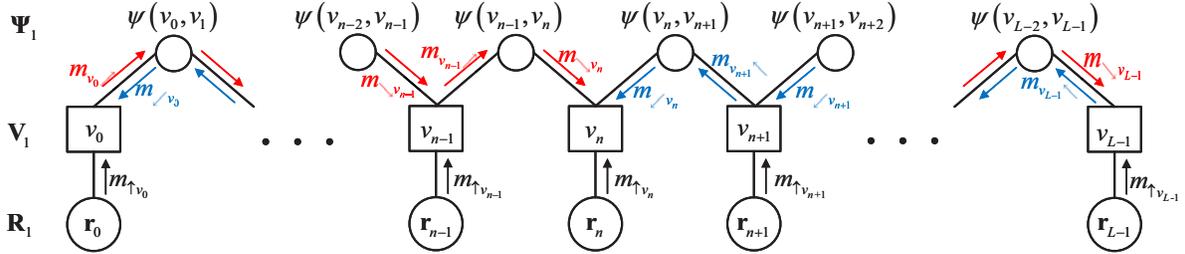}\\
  \caption{Graphical interpretation of \eqref{eq:III-B1} giving rise to a BP algorithm on a Tanner graph of the first block of \textbf{R}.}\label{Fig4}
\end{figure}

Let us interpret \eqref{eq:III-B1} in the context of a BP message passing algorithm on the Tanner graph with the aid of Fig.\ref{Fig4}. For simplicity, we look at the first block of ${\bf{R}}$. In Fig.\ref{Fig4}, ${{\bf{R}}_{\rm{1}}}$ denotes the evidence nodes; ${{\bf{V}}_{\rm{1}}}{\rm{ = }}\left( {{v_{\rm{0}}}, \cdots ,{v_{L - 1}}} \right)$ denotes the variable nodes, where ${v_n} = \left( {{s_n},{{\tilde \theta }_n}} \right)$. Note that ${s_n}$ is a discrete variable and ${\tilde \theta _n}$ is a continuous variable; and ${{\bf{\Psi }}_{\rm{1}}}$  denotes the check nodes. The relationship between two adjacent variables nodes ${v_n}$ and ${v_{n + 1}}$ is formulated as
\begin{align}
{\psi _{\rm{1}}}\left( {{v_n},{v_{n + 1}}} \right) = \delta \left( {{{\tilde \theta }_{n + 1}} - {{[{{\tilde \theta }_n} + \vartheta ]}_{2\pi }}} \right)
\end{align}
The BP algorithm is summarized below:
\begin{enumerate}[(i)]
\item Compute upward messages.\\
      First, we compute the messages $\Pr\left( {{{\bf{r}}_n}\left| {{s_n},{{\tilde \theta }_n}} \right.} \right) = \Pr\left( {{{\bf{r}}_n}\left| {{v_n}} \right.} \right)$ generated by the observations ${{\bf{r}}_n}$ for $n \in \left[ {{\rm{0}},L - 1} \right]$.  The expression of $\Pr\left( {{{\bf{r}}_n}\left| {{v_n}} \right.} \right)$ is derived in Appendix D\footnote{Appendix D derives an expression for general channel gains  $\left| {{h_A}} \right|$ and $\left| {{h_B}} \right|$. We draw on the results from Appendix D in the special case in this subsection, where $\left| {{h_A}} \right| = \left| {{h_B}} \right|{\rm{ = 1}}$.}. We denote $\Pr\left( {{{\bf{r}}_n}\left| {{v_n}} \right.} \right)$  by ${m_{ \uparrow {v_n}}}$ in Fig.\ref{Fig4}.
\item Compute right-bound messages.\\
      The right-bound messages are computed from the leftmost part to the rightmost part of the Tanner graph. In Fig.\ref{Fig4}, ${m_{{v_{n - 1}} \nearrow }}$ is the message from node ${v_{n - 1}}$ to node $\psi \left( {{v_{n - 1}},{v_n}} \right)$ and ${m_{ \searrow {v_n}}}$  is the message from node $\psi \left( {{v_{n - 1}},{v_n}} \right)$ to node ${v_n}$, $n \in \left[ {1,L - 1} \right]$.  The right-bound messages are computed as follows:
      \begin{align}
      {m_{{v_{n - 1}} \nearrow }}&={m_{ \searrow {v_{n - 1}}}}{m_{ \uparrow {v_{n - 1}}}}\\
      {m_{ \searrow {v_n}}}&=\int d {{\tilde \theta }_{n - 1}}\sum\limits_{{s_{n - 1}}} {{m_{{v_{n - 1}} \nearrow }}} \psi \left( {{v_{n{\rm{ - 1}}}},{v_n}} \right)
      \end{align}
\item Compute left-bound messages. \\
      The left-bound messages are computed from the rightmost part to the leftmost part of the Tanner graph. In Fig.\ref{Fig4}, ${m_{{v_{n + 1}} \nwarrow }}$ is message from node ${v_{n + 1}}$ to node $\psi \left( {{v_n},{v_{n + 1}}} \right)$ and ${m_{ \swarrow {v_n}}}$ is the message from node $\psi \left( {{v_n},{v_{n + 1}}} \right)$ to node ${v_n}$. The left-bound messages are computed as follows:
       \begin{align}
       {m_{{v_{n + 1}} \nwarrow }}&={m_{ \swarrow {v_{n + 1}}}}{m_{ \uparrow {v_{n + 1}}}}\\
       {m_{ \swarrow {v_n}}}&=\int d {{\tilde \theta }_{n + 1}}\sum\limits_{{s_{n + 1}}} {{m_{{v_{n + 1}} \nwarrow }}} \psi \left( {{v_n},{v_{n{\rm{ + 1}}}}} \right)
       \end{align}
\item Compute a posteriori probabilities.\\
      The tanner graph in Fig.\ref{Fig4} has a tree structure. Therefore, $\Pr\left( {\left. {{s_n},{{\tilde \theta }_n},\vartheta } \right|{{\bf{R}}_{\rm{1}}}} \right)$ can be computed exactly through one iteration. Specifically,
      \begin{align}
      \Pr\left( {\left. {{s_n},{{\tilde \theta }_n},\vartheta } \right|{{\bf{R}}_{\rm{1}}}} \right) = {m_{ \uparrow {v_n}}}{m_{ \searrow {v_n}}}{m_{ \swarrow {v_n}}}
      \end{align}
\end{enumerate}

Once $\Pr\left( {\left. {{s_n},{{\tilde \theta }_n},\vartheta } \right|{{\bf{R}}_q}} \right)$ for $q \in \left[ {1,Q} \right]$ are computed by the BP algorithm, we can get the $\Pr\left( {\left. {{s_n}} \right|{{\bf{R}}_q}} \right)$ by marginalizing over ${\tilde \theta _n}$ and $\vartheta $. The detector then makes the decision
\begin{align}
s_n^*{\rm{ = }}\mathop {{\rm{arg max }}}\limits_{{s_n} = 0,1}\Pr\left( {\left. {{s_n}} \right|{{\bf{R}}_q}} \right)
\end{align}
in the \emph{q}-th block. In this paper, we refer to this detector as \textbf{Brief Propagation Detector (BPD)}.

The complexity of BPD is in the order of the packet length \emph{N}, i.e., ${\rm O}\left( N \right)$. The complexity of MPD is also ${\rm O}\left( N \right)$.

\subsection{Numerical Results }
We now evaluate the BER performance of BPD numerically. We benchmark BPD against two schemes: 1) MPD; 2) PerfPD: a detector that has perfect knowledge of the relative phase. Comparison of BPD with MPD shows that by leveraging the relationship of the relative phase among successive received-signal-magnitudes, BPD can have significantly
better BER than MPD. Furthermore, comparison of BPD with PerfPD shows that BPD can have
BER performance approaching that of PerfPD.

For the numerical study, we assume the packets are 128 bits (symbols) in length, the CFO $\tilde f \in \left[ { - 10{\kern 1pt} {\kern 1pt} {\rm{kHz}},10{\kern 1pt} {\kern 1pt} {\rm{kHz}}} \right]$, and the symbol duration is $T=1$ $\mu s$. In addition, BPD, MPD, and PerfPD have \emph{a priori} knowledge of the  channel gains $|{h_A}| = |{h_B}| = 1$, but not the relative phases.

\begin{figure}[ht]
  \centering
  \includegraphics[width=0.6\columnwidth]{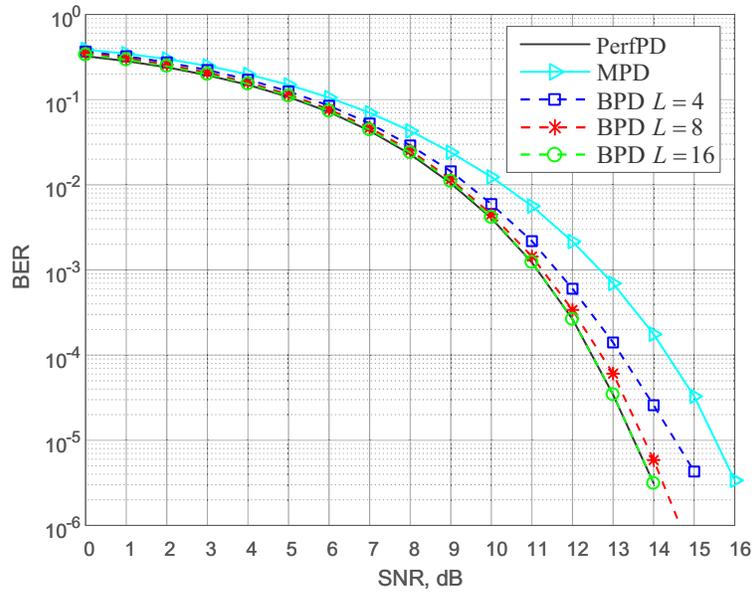}\\
  \caption{BER of PerfPD, MPD, and BPD with different block lengths, when the relative phase is fixed to ${\rm{0}}{\rm{.2}}\pi $  for the whole packet.}\label{Fig5}
\end{figure}

\begin{figure}[ht]
  \centering
  \includegraphics[width=0.6\columnwidth]{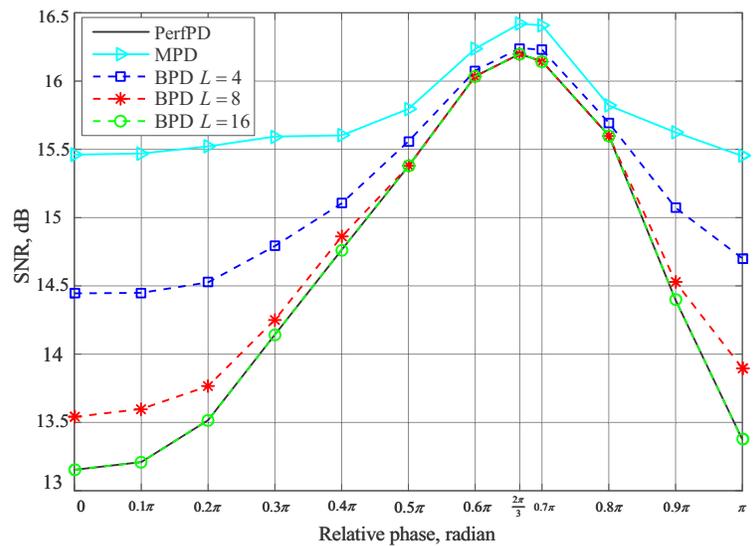}\\
  \caption{SNR (in dB) required for ${\rm{BER = 1}}{{\rm{0}}^{{\rm{ - 5}}}}$ versus relative phase for PerfPD, MPD, and BPD with different block lengths. Relative phase is assumed to be constant within a packet.  }\label{Fig6}
\end{figure}

\begin{figure}[h]
  \centering
  \includegraphics[width=0.6\columnwidth]{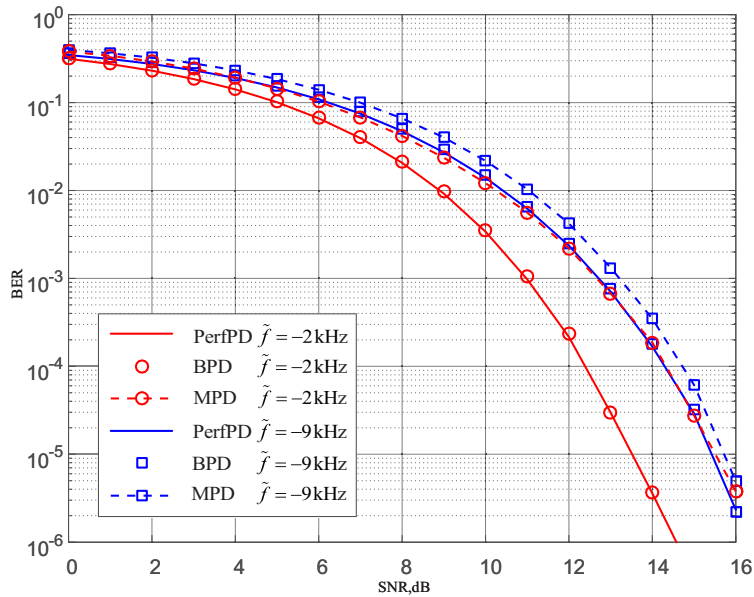}\\
  \caption{BER versus SNR for PerfPD, BPD, and MPD. The initial relative phase is ${\rm{0}}{\rm{.2}}\pi $. CFOs of -2 kHz and -9 kHz  are examined. The block length of BPD is \emph{L}=16.}\label{Fig7}
\end{figure}

Fig.\ref{Fig5} and Fig.\ref{Fig6} present the results of the case in which the relative phase stays constant within each overall packet, i.e., no CFO between users \emph{A} and \emph{B}. Fig.\ref{Fig7} presents the results of the case in which the relative phase changes incrementally from symbol to symbol due to a CFO between users \emph{A} and \emph{B}. In this case, BPD estimates both the initial relative phase and the CFO.

In Fig.\ref{Fig5}, the relative phase ${\tilde \theta _n}$ is set to ${\rm{0}}{\rm{.2}}\pi $ for all \emph{n}. The dashed lines are the BER results of BPD with block lengths $L = 4, 8, 16$. The solid line with triangle markers is the BER of MPD. The black line is the BER of PerfPD. Compared with MPD, BPD with $L=4$ and $L=16$ have 1.00 dB and 2.01 dB performance improvements at ${\rm{BER = 1}}{{\rm{0}}^{{\rm{ - 5}}}}$. The performance gap between BPD and PerfPD is small. For example, for BPD with $L=8$, the performance gap with the idealized PerfPD is only 0.25 dB at ${\rm{BER = 1}}{{\rm{0}}^{{\rm{ - 5}}}}$. BPD with $L=16$ performs as well as PerfPD with no gap. This phenomenon suggests that, for optimal performance of BPD, the relative phase only needs to be constant within 16 successive symbols (so that we could use 16 successive symbols for each block in our BP algorithm), although earlier in the paper, we made the conservative assumption that the whole packets are phase coherent.

Fig.\ref{Fig6} plots the SNR required for the various detectors to achieve BER of ${\rm{1}}{{\rm{0}}^{{\rm{ - 5}}}}$. Again, the relative phase is constant within each packet. We plot the curves between $\left[ {0,\pi } \right]$ only, since the BERs of ${\tilde \theta _n}$  and $ - {\tilde \theta _n}$  are the same (see Appendix D). Fig.\ref{Fig6} shows that the required SNR changes as the relative phase varies. This validates an important aspect of noncoherent PNC detection pointed out by our paper. Specifically, unlike a noncoherent detector for point-to-point communication, the performance of a noncoherent PNC detector is affected by the relative phase. From Fig.\ref{Fig6}, benchmarked against PerfPD, BPD with block length $L=16$ is sufficient for optimal performance for all relative phases. Another point to note is that the performance gap between BPD with $L=16$ and MPD changes with relative phase. In general, we have a maximum improvement of 2.31 dB at the relative phase ${\tilde \theta _n}{\rm{ = 0}}$.

Fig.\ref{Fig7} shows BER versus SNR for the various detectors with an initial relative phase of  $0.2\pi$ and constant CFOs of -2 kHz (red lines) and -9 kHz (blue lines).

For the red lines, with CFO of -2 kHz, the relative phase ${\tilde \theta _n}$ changes from symbol to symbol with a constant decrement of $0.004\pi $ per symbol. Thus, the relative phase ranges from $0.2\pi $ at the beginning of the packet to $-0.312\pi $ at the end of the packet. At ${\rm{BER = 1}}{{\rm{0}}^{{\rm{ - 5}}}}$, the performance gap between BPD with $L=16$ and the MPD is 2.01 dB; and BPD with $L=16$ performs as well as PerfPD with no gap. Again, as in Fig.\ref{Fig5} and Fig.\ref{Fig6}, BPD with $L=16$ is optimal even if there is a CFO.

For the blue lines, with CFO of -9 kHz, the relative phase ${\tilde \theta _n}$ changes from $0.2\pi $  at the beginning of the packet to $-2.104\pi$  at the end of the packet. Thus, from Fig.\ref{Fig6}, we see that the relative phase within the packet of this set-up covers the whole spectrum of the cases, from significant performance gaps to small performance gaps between BPD (or PerfPD) and MPD. At ${\rm{BER = 1}}{{\rm{0}}^{{\rm{ - 5}}}}$, the performance gap between BPD with $L=16$ and the MPD is 0.30 dB; and BPD with $L=16$ performs as well as PerfPD with no gap.

Fig.\ref{Fig7} shows that BPD in general has BER improvements compared with MPD, and BPD with $L=16$ performs optimally when benchmarked against PerfPD. Although Fig.\ref{Fig7} only shows two cases of CFOs only, we believe that they are representative. Specifically, as shown in Fig.\ref{Fig6}, in the case of $\tilde f =  - 2{\kern 1pt} {\kern 1pt} {\rm{kHz}}$, the relative phases within the packet lie within the range in which the performance gap between BPD (or PerfPD) and MPD is significant. On the other hand, in the case of $\tilde f =  - 9{\kern 1pt} {\kern 1pt} {\rm{kHz}}$, the relative phases within the packet cover the whole spectrum of performance gaps between BPD (or PerfPD) and MPD.

\subsection{Power-Imbalanced Channels}
Subsections IV-A, IV-B, and IV-C assumed power-balanced channels, and that channel gains $\left| {{h_A}} \right|{\rm{ = }}\left| {{h_{\rm{B}}}} \right|$ were perfectly known at BPD. To explicitly denote the fact that we know the channel gains to be $\left| {{h_A}} \right|$ and $\left| {{h_B}} \right|$ in the \emph{q}-th block, the integrand in \eqref{eq:III-B00} in Subsection IV-B can be rewritten as
\begin{align}
\Pr\left( {\left. {{s_n},{{\tilde \theta }_n},\vartheta } \right|{{\bf{R}}_q}} \right)=\Pr\left( {\left. {{s_n},{{\tilde \theta }_n},\vartheta } \right|{{\bf{R}}_q},\left| {{h_A}} \right|{\rm{,}}\left| {{h_B}} \right|} \right)  \label{eq:V-2}
\end{align}
for $n = (q - 1)L, \cdots ,qL - 1$.

This subsection relaxes the assumption of $\left| {{h_A}} \right| = \left| {{h_B}} \right|$ and consider power-imbalanced channels. The expression in \eqref{eq:V-2} remains valid; just that  $\left| {{h_A}} \right|$ and $\left| {{h_B}} \right|$  are not necessarily equal.

The BP algorithm as expounded in Subsection IV-B can be easily generalized to the power-imbalanced and known-channel-gains set-up. In particular, the algorithm in Subsection IV-B builds on the more general results of Appendix C, which gives the BP recursive breakdown of \eqref{eq:V-2} as
\begin{equation}
\begin{split}
&\Pr\left( {\left. {{s_n},{{\tilde \theta }_n},\vartheta } \right|{{\bf{R}}_q},\left| {{h_A}} \right|,\left| {{h_B}} \right|} \right)\\
&= {\eta _q}\Pr\left( \vartheta  \right)\Pr\left( {{{\bf{r}}_n}\left| {{s_n},{{\tilde \theta }_n},\left| {{h_A}} \right|,\left| {{h_B}} \right|} \right.} \right)\\
&\times \int {d{{\tilde \theta }_{n - 1}} \cdots d{{\tilde \theta }_{\left( {q - 1} \right)L}}} \prod\limits_{i = (q - 1)L}^{n - 1} {\sum\limits_{{s_i}} {\Pr\left( {{{\bf{r}}_i}\left| {{s_i},{{\tilde \theta }_i},\left| {{h_A}} \right|,\left| {{h_B}} \right|} \right.} \right)\delta \left( {{{\tilde \theta }_i} - {{\left[ {{{\tilde \theta }_{i + 1}} - \vartheta } \right]}_{{\rm{2}}\pi }}} \right)} }\\
&\times \int {d{{\tilde \theta }_{n + 1}} \cdots d{{\tilde \theta }_{qL - 1}}} \prod\limits_{i = n + 1}^{qL - 1} {\sum\limits_{{s_i}} {\Pr\left( {{{\bf{r}}_i}\left| {{s_i},{{\tilde \theta }_i},\left| {{h_A}} \right|,\left| {{h_B}} \right|} \right.} \right)\delta \left( {{{\tilde \theta }_i} - {{\left[ {{{\tilde \theta }_{i - 1}} + \vartheta } \right]}_{{\rm{2}}\pi }}} \right)} }
\end{split}  \label{eq:V-3}
\end{equation}
That is, the generalization has already been given in Appendix C. Using this more general set-up, the BP algorithm remains the same as that described in Subsection IV-B; just that we set $\left| {{h_A}} \right|$ and $\left| {{h_B}} \right|$ to be general values in the BP algorithm.

\section{Detector Design with Unknown Channel Gains}
\begin{figure}[ht]
  \centering
  \includegraphics[scale=0.90]{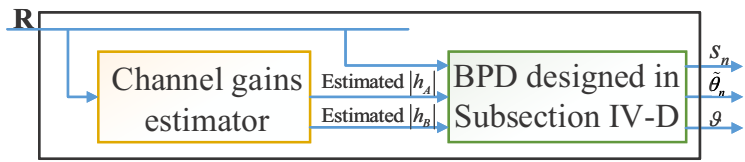}\\
  \caption{Overall detector with the magnitudes of received signals \textbf{R} being the input, and ${s_n}$, ${\tilde \theta _n}$, and  $\vartheta $ being the outputs for $n = {\rm{0}}, \cdots ,N - {\rm{1}}$.}\label{Fig8}
\end{figure}

This section considers the most general set-up with power-imbalanced channels and unknown channel gains. Since we do not know $\left| {{h_A}} \right|$ and $\left| {{h_B}} \right|$ beforehand, the detector in Subsection IV-D cannot be applied directly. Our approach is to add a channel gains estimator for $\left| {{h_A}} \right|$ and $\left| {{h_B}} \right|$ so that we can feed the estimated $\left| {{h_A}} \right|$ and $\left| {{h_B}} \right|$ to the BPD in Subsection IV-D. The new framework is shown in Fig.\ref{Fig8}. The newly added channel gains estimator is the first block on the left. The BPD in Subsection IV-D corresponds to the second block of Fig.\ref{Fig8}, for which the magnitudes of received signals \textbf{R} , and the estimated channel gains $\left| {{h_A}} \right|$ and $\left| {{h_B}} \right|$ from the first block, are the inputs, and ${s_n}$, ${\tilde \theta _n}$, and $\vartheta $ are the outputs.

This overall detector, as shown in Fig.\ref{Fig8}, is applicable to scenarios with block fading channels: the channel gains $\left| {{h_A}} \right|$ and $\left| {{h_B}} \right|$ are constant within each packet but may vary from packet to packet. For the block fading scenario, the BER is a weighted average of the BER of set-ups with different instances of $\left| {{h_A}} \right|$ and $\left| {{h_B}} \right|$. Subsection V-C will present numerical results for our overall detector when applied to block Rayleigh fading channels.

\begin{figure}[ht]
  \centering
  \includegraphics[scale=0.90]{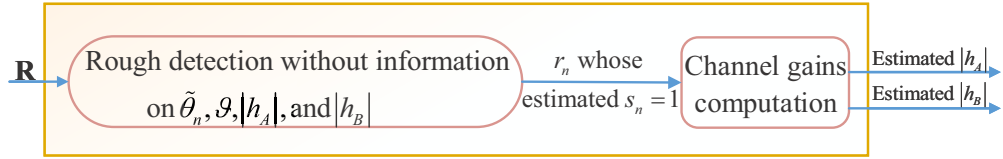}\\
  \caption{Processes within the channel gains estimator (the first block of Fig.\ref{Fig8}).}\label{Fig9}
\end{figure}

Continuing with Fig.\ref{Fig8}, recall that we do not have preambles in our set-up. Therefore, the channel gains estimator in the first block has to count on the data (the whole packet is the data portion), \textbf{R}, for the estimation of $\left| {{h_A}} \right|$ and $\left| {{h_B}} \right|$. An issue, however, is that in the received data symbols \textbf{R}, unknown variables ${s_n}$, ${\tilde \theta _n}$, $\vartheta $, $\left| {{h_A}} \right|$ and $\left| {{h_B}} \right|$ are intertwined with each other. Fortunately, when users \emph{A} and \emph{B} transmit on different frequencies (when ${s_n} = 1$), the uplink PNC can be viewed as two parallel point-to-point communication channels on two different frequencies with non-overlapping signals from users \emph{A} and \emph{B}. The signal magnitudes on the two different frequencies contain the information of $\left| {{h_A}} \right|$ and $\left| {{h_B}} \right|$ separately, and are not affected by the phase. In other words, when ${s_n} = 1$, we can estimate the channel gains $\left| {{h_A}} \right|$ and $\left| {{h_B}} \right|$ directly from the magnitude and do not need to know the relative phase and CFO. Our design of the channel gains estimator, as shown in Fig.\ref{Fig9}, is to first attempt to isolate the symbols for which ${s_n} = 1$ (the first block in Fig.\ref{Fig9}), and then use these isolated symbols to estimate $\left| {{h_A}} \right|$ and $\left| {{h_B}} \right|$ (the second block in Fig.\ref{Fig9}).

To isolate the symbols for which ${s_n} = 1$, the received signal magnitudes \textbf{R} go through a rough detection process first (we may make mistakes in identifying symbols for which ${s_n} = 1$). This rough detection process does not have the phase and channel gains information. The phase estimation, together with fine detection of ${s_n}$, will be performed by the BPD (the second block in Fig.\ref{Fig8}) after $\left| {{h_A}} \right|$ and $\left| {{h_B}} \right|$ are estimated by the channel gains computation in the second block of Fig.\ref{Fig9}.

In the following, we will describe the details of the rough detection process and the channel-gain computation process in Fig.\ref{Fig9}.

\subsection{K-means Clustering Detection}
For the first block in Fig.\ref{Fig9}, we use a K-means clustering detector to group the data samples ${{\bf{r}}_n}$ into two types: those for which ${s_n} = 1$ and those for which ${s_n} = 0$. When ${s_n} = 1$, \emph{A} and \emph{B} transmit on the different frequencies, and therefore both frequencies contain signals and noise. On the other hands, when ${s_n} = 0$, \emph{A} and \emph{B} transmit on the same frequency. Without loss of generality, suppose that they both transmit on the first frequency. Then the first frequency contains superimposed signals of \emph{A} and \emph{B}, and noise; and the second frequency contains only noise. In this case, we have $\Pr\left( {\left| {r_{n,1}^{}} \right| > \left| {r_{n,2}^{}} \right|} \right)\gg\Pr\left( {\left| {r_{n,1}^{}} \right| \le \left| {r_{n,2}^{}} \right|} \right)$. The difference between the two scenarios can be leveraged to differentiate the two types of signals. Specifically, let ${\hat r_n} = \min \left( {\left| {r_{n,1}^{}} \right|,\left| {r_{n,2}^{}} \right|} \right)$, so that when ${s_n} = 1$, ${\hat r_n}$ contains signal and noise; and when ${s_n} = 0$, in most cases, ${\hat r_n}$ only contains noise. Based on ${\hat r_n}$, we partition the magnitudes of received signals ${{\bf{r}}_n} = \left( {\left| {{r_{n,1}}} \right|,\left| {{r_{n,2}}} \right|} \right)$ into two clusters
\begin{align}
{I_0} &= \left\{{{\bf{r}}_n}: {{\bf{r}}_n} \text{ is grouped under the XOR=0 cluster}\right\}\\
{I_1} &= \left\{{{\bf{r}}_n}: {{\bf{r}}_n} \text{ is grouped under the XOR=1 cluster}\right\}
\end{align}
such that
\begin{align}
\sum\limits_{i = 0}^1 {\sum\limits_{{{\hat r}_n}{\kern 1pt} {\rm{for}}{\kern 1pt} {\kern 1pt} {{\bf{r}}_n} \in {I_i}} {{{\left\| {{{\hat r}_n} - {{\bar r}_i}} \right\|}^2}} }
\end{align}
is minimized, where ${\bar r_i}$ is the mean of points ${\hat r_n}$ in the cluster ${I_i}$. The partitions ${I_0}$  and ${I_1}$   can be found by the K-means clustering algorithm\cite{ mackay2003information }.

Before running the algorithm, we need to set initial values of ${\bar r_0}$ and ${\bar r_1}$. Since ${\hat r_n}$ in $I_1$ contain both signal and noise, the mean of points ${\hat r_n}$ in ${I_1}$ should be larger than that in the first cluster ${I_0}$, i.e., ${\bar r_0} < {\bar r_1}$. We set the minimum of all ${\hat r_n}$  as the initial value of ${\bar r_0}$, i.e., ${\bar r_0}^{\left( 0 \right)} = \min ({\{ {\hat r_n}\} _{n = 0, \cdots ,N - 1}})$, and the maximum of all ${\hat r_n}$ as the initial value of ${\bar r_1}$, i.e., ${\bar r_1}^{\left( 0 \right)} = \max ({\{ {\hat r_n}\} _{n = 0, \cdots ,N - 1}})$.

From the received signal ${{\bf{r}}_n} = \left( {\left| {{r_{n,1}}} \right|,\left| {{r_{n,2}}} \right|} \right)$, we find ${\hat r_n} = \min \left( {\left| {r_{n,1}^{}} \right|,\left| {r_{n,2}^{}} \right|} \right)$, $n = {\rm{0}},\cdots,N{\rm{ - 1}}$. The algorithm proceeds by alternating between the following two steps:
\begin{enumerate}[(i)]
\item \textbf{Assignment step}: In the \emph{t}-th iteration, assign each  ${{\bf{r}}_n}$ to the cluster whose mean has the least squared Euclidean distance to ${\hat r_n}$  according to the following equations:
    \begin{align}
    {I_0} &= \left\{ {{{\bf{r}}_n}:{{\left\| {{{\hat r}_n} - {{\bar r}_0}^{\left( t \right)}} \right\|}^2} \le {{\left\| {{{\hat r}_n} - {{\bar r}_1}^{\left( t \right)}} \right\|}^2}} \right\}\\
    {I_1} &= \left\{ {{{\bf{r}}_n}:{{\left\| {{{\hat r}_n} - {{\bar r}_1}^{\left( t \right)}} \right\|}^2} < {{\left\| {{{\hat r}_n} - {{\bar r}_0}^{\left( t \right)}} \right\|}^2}} \right\}
    \end{align}
    where ${\bar r_i}^{\left( t \right)}$ is the mean of points ${\hat r_n}$  in the cluster ${I_i}$  in the \emph{t}-th iteration.
\item \textbf{Update step}: For ${{\bf{r}}_n} \in {I_0}$, update the mean of the cluster ${I_0}$  by ${\bar r_0}^{\left( {t + 1} \right)} ={\sum {{{\hat r}_n}} }/{\left| {{I_0}} \right|}$; for ${{\bf{r}}_n} \in {I_1}$, update the mean of the cluster ${I_1}$  by ${\bar r_1}^{\left( {t + 1} \right)} ={\sum {{{\hat r}_n}} }/{\left| {{I_1}} \right|}$.
\end{enumerate}

The iterations are stopped when  ${I_0}$ and  ${I_1}$ do not change anymore. In this paper, we refer to this detector as \textbf{K-means clustering detector (KD)}. After KD, the points in ${I_1}$ are used in the second block of Fig.\ref{Fig9} for the computation of the channel gains.

\subsection{Channel Gains Computation}
Define $\left| {{h_{\min }}} \right|{\rm{ = }}\min \left( {\left| {{h_A}} \right|,\left| {{h_B}} \right|} \right)$  and $\left| {{h_{\max }}} \right|{\rm{ = }}\max \left( {\left| {{h_A}} \right|,\left| {{h_B}} \right|} \right)$. With the signal ${{\bf{r}}_n} = \left( {\left| {{r_{n,1}}} \right|,\left| {{r_{n,2}}} \right|} \right)$  belonging to ${I_1}$ , the procedures of the channel gains computation algorithm in the second block of Fig.\ref{Fig9} are summarized as follows:
\begin{enumerate}[(i)]
\item Get rough computations of $\left| {{h_{\min }}} \right|$  and $\left| {{h_{\max }}} \right|$  as follows:
      \begin{align}
      {\left| {\hat h_{\min }^{{\rm{rough}}}} \right|^{\rm{2}}} ={\sum\limits_{{{\bf{r}}_n} \in {I_1}}^{} {\min \left( {{{\left| {r_{n,1}^{}} \right|}^{\rm{2}}},{{\left| {r_{n,2}^{}} \right|}^{\rm{2}}}} \right)} }/{\left| {{I_1}} \right|} - {N_0}
      \end{align}
      and
      \begin{align}
      {\left| {\hat h_{\max }^{{\rm{rough}}}} \right|^{\rm{2}}} ={\sum\limits_{{{\bf{r}}_n} \in {I_1}}^{} {\max \left( {{{\left| {r_{n,1}^{}} \right|}^{\rm{2}}},{{\left| {r_{n,2}^{}} \right|}^{\rm{2}}}} \right)} }/{\left| {{I_1}} \right|} - {N_0}
      \end{align}
      where ${N_0}$  is a bias when we use the square of magnitudes to compute channel gains.
\item Perform fine computations of $\left| {{h_{\min }}} \right|$  and $\left| {{h_{\max }}} \right|$ by searching over the intervals
     \begin{align}
     {\Omega _{\min }} = \left[ {\max \left( {\left| {\hat h_{\min }^{{\rm{rough}}}} \right| - \beta ,0} \right),\left| {\hat h_{\min }^{{\rm{rough}}}} \right| + \beta } \right]
     \end{align}
     and
     \begin{align}
     {\Omega _{\max }} = \left[ {\max \left( {\left| {\hat h_{\max }^{{\rm{rough}}}} \right| - \beta ,0} \right),\left| {\hat h_{\max }^{{\rm{rough}}}} \right| + \beta } \right]
     \end{align}
     respectively, for some interval-length parameter $\beta  > 0$. Find
    \begin{align}
    \left( {\left| {\hat h_{\min }^{{\rm{fine}}}} \right|,\left| {\hat h_{\max }^{{\rm{fine}}}} \right|} \right) = \mathop {\arg\max }\limits_{\left| {{h_{\min }}} \right| \in {\Omega _{\min }},{\kern 1pt} \left| {{h_{\max }}} \right| \in {\Omega _{\max }}} \prod\limits_{{{\bf{r}}_n} \in {I_1}} {\Pr\left( {\left. {{{\bf{r}}_n}} \right|{s_n} = 1,\left| {{h_{\min }}} \right|,\left| {{h_{\max }}} \right|} \right)}
    \end{align}
    where $\Pr\left( {\left. {{{\bf{r}}_n}} \right|{s_n} = 1,\left| {{h_{\min }}} \right|,\left| {{h_{\max }}} \right|} \right)$  is given in Appendix D; $\left| {\hat h_{\min }^{{\rm{fine}}}} \right|$  is the estimated channel-gain of $\left| {{h_{\min }}} \right|$ , and $\left| {\hat h_{\max }^{{\rm{fine}}}} \right|$  is the estimated channel-gain of $\left| {{h_{\max }}} \right|$.
\end{enumerate}

After executing the two processes described above, we have the estimated channel gains $\left| {\hat h_{\min }^{{\rm{fine}}}} \right|$  and $\left| {\hat h_{\max }^{{\rm{fine}}}} \right|$. The channel gains estimator in the first block of Fig.\ref{Fig8} feeds the estimated channel gains $\left| {\hat h_{\min }^{{\rm{fine}}}} \right|$  and $\left| {\hat h_{\max }^{{\rm{fine}}}} \right|$ to BPD in the second block of Fig.\ref{Fig8}. Given this knowledge, BPD in turn jointly detects ${s_n}$ , ${\tilde \theta _n}$, and $\vartheta $. Since we use KD for rough detection in the first process of the channel gains estimator and BPD for fine detection, we refer to this overall detector shown in Fig.\ref{Fig8} as \textbf{KD-BPD}.

\subsection{Numerical Results}
\begin{figure}[ht]
  \centering
  \subfigure[]
  {
  \label{Fig10a}
  \includegraphics[width=0.46\columnwidth]{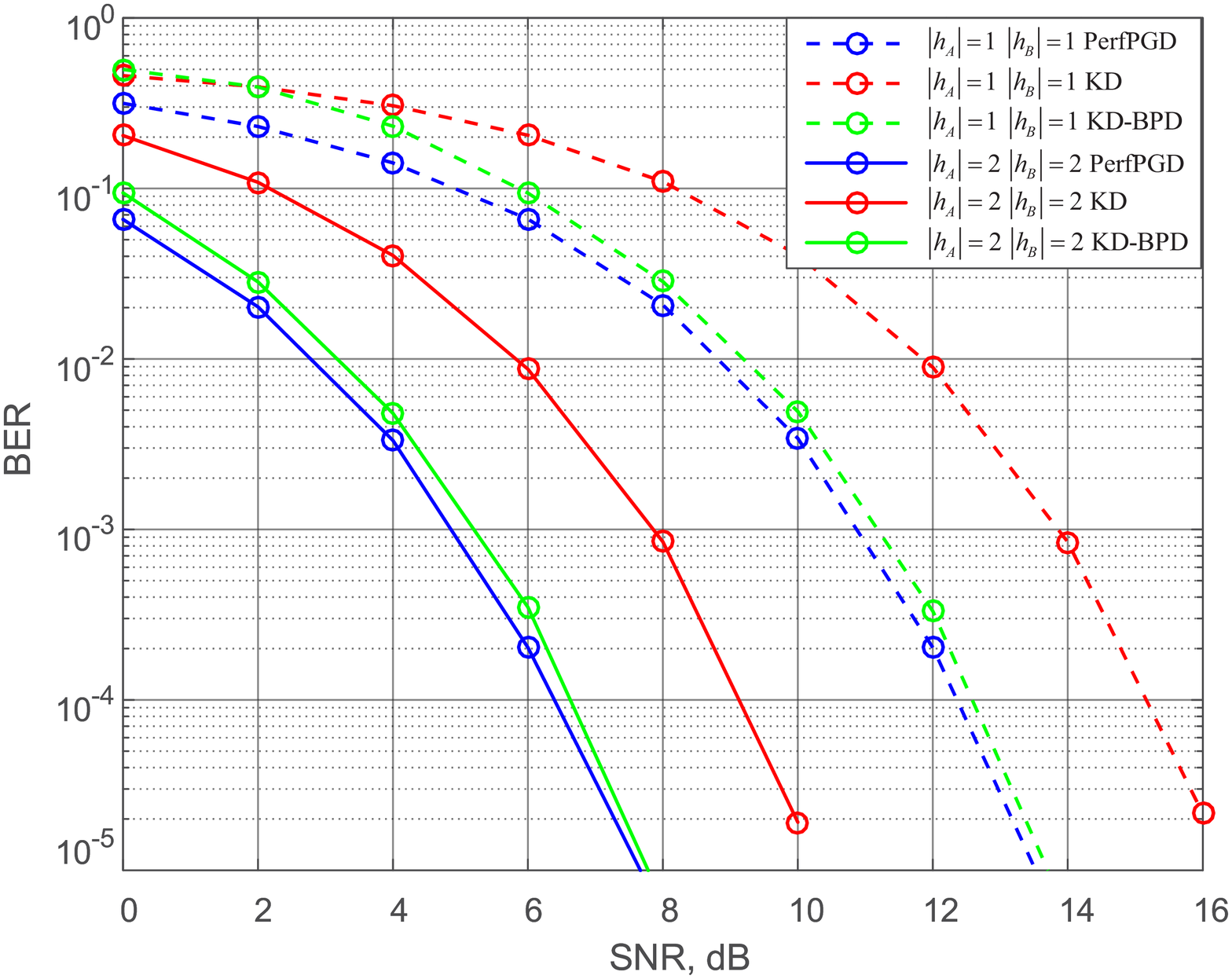}
  }
  \subfigure[]
  {
  \label{Fig10b}
  \includegraphics[width=0.46\columnwidth]{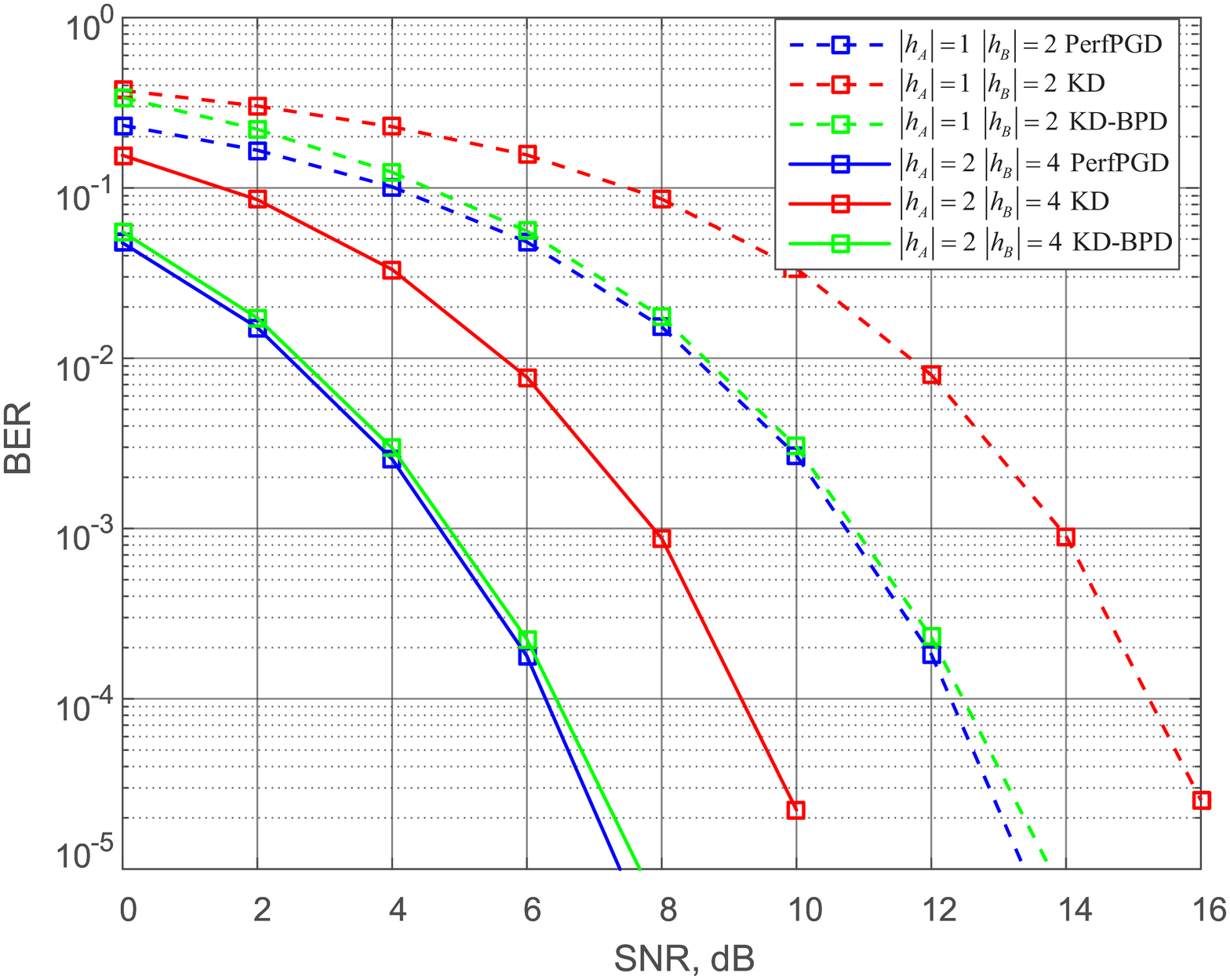}
  }
  \subfigure[]
  {
  \label{Fig10c}
  \includegraphics[width=0.46\columnwidth]{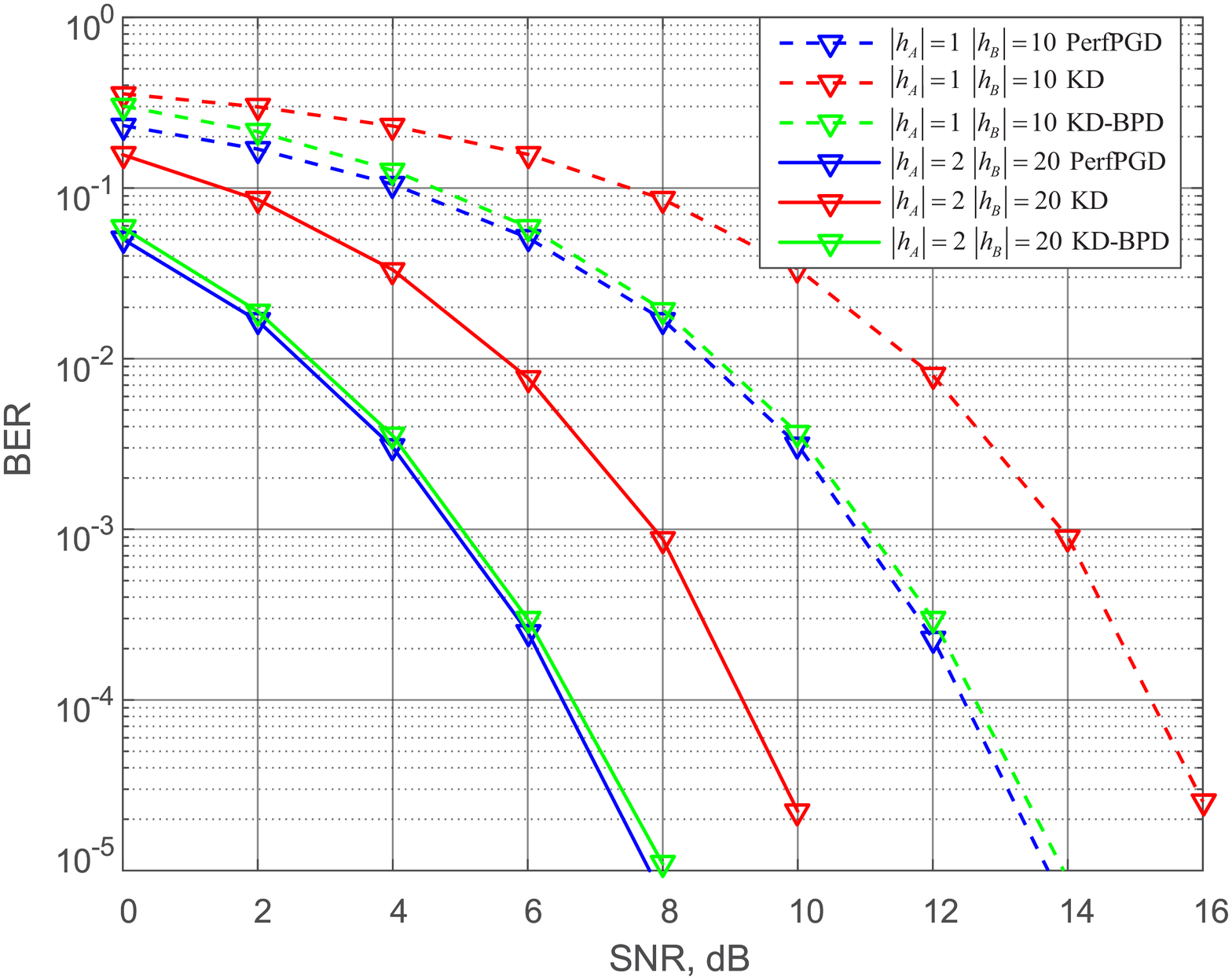}
  }
  \caption{BER of PerfPGD, KD, and KD-BPD in different power-imbalanced channels. The block length in BPD is  $L=16$. The initial relative phase is ${\rm{0}}{\rm{.2}}\pi$  and CFO $\tilde f =  - 2{\kern 1pt} {\kern 1pt} {\rm{kHz}}$ between \emph{A} and \emph{B}. The SNR on X-axis is defined as ${\rm{SNR = 10lo}}{{\rm{g}}_{10}}\left( {{E_s^A}/{{N_0}}} \right){\rm{ = 10lo}}{{\rm{g}}_{10}}\left( {{E_s^B}/{{N_0}}} \right)$, where  $E_s^A$ and $E_s^B$ are the transmitted signal power from users \emph{A} and \emph{B} respectively, and  $E_s^A = E_s^B$.}\label{Fig10}
\end{figure}

\begin{figure}[ht]
  \centering
  \includegraphics[width=0.6\columnwidth]{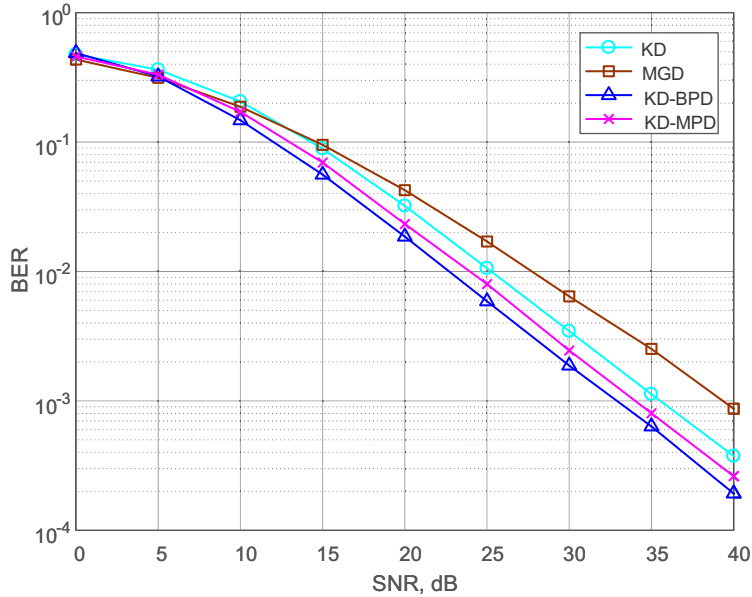}\\
  \caption{BER of KD, MGD, KD-BPD, and KD-MPD when the initial relative phase is ${\rm{0}}{\rm{.2}}\pi $, CFO $\tilde f =  - 2{\kern 1pt} {\kern 1pt} {\rm{kHz}}$ between \emph{A} and \emph{B}. We compare these detectors in an average-power-balanced Rayleigh fading channels and assume $E\left( {{{\left| {{h_A}} \right|}^2}} \right) = E\left( {{{\left| {{h_B}} \right|}^2}} \right) = 1$. The block length in BPD is $L=16$.}\label{Fig11}
\end{figure}

We assume the packets are 128 bits in size, CFO $\tilde f \in \left[ { - 10{\kern 1pt} {\kern 1pt} {\rm{kHz}},10{\kern 1pt} {\kern 1pt} {\kern 1pt} {\rm{kHz}}} \right]$ and the symbol duration is $T=1$ $\mu s$. Fig.\ref{Fig10} studies the case where the channels are power-imbalanced and channel gains $\left| {{h_A}} \right|$   and $\left| {{h_B}} \right|$  are constant among different packets; Fig.\ref{Fig11} studies the case where the channel gains are Rayleigh distributed among different packets. The definitions of detectors under testing are summarized as follows:
\begin{enumerate}
\item \textbf{KD:} A detector that detects symbols by applying the K-means clustering algorithm. This detector only consists of the first block in Fig.\ref{Fig9} for the rough detection. This detector is designed by us.
\item \textbf{KD-BPD:} A detector that applies KD for rough symbol detection, then computes channel gains based on the rough detection, and then feeds the estimated channel gains to BPD for fine symbol detection. The overall scheme follows that of Fig.\ref{Fig8}. This detector is designed by us.
\item \textbf{PerfPGD:} A detector that has perfect knowledge of phases and channel gains. This ideal detector serves as a benchmark in this paper.
\item \textbf{MGD} (Marginalized-channel-gain Detector)\textbf{:} A detector, proposed in \cite{valenti2011noncoherent}, that detects signals by approximating the relative phase to be $\pi/2$  and marginalizing the channel gains $\left| {{h_A}} \right|$  and $\left| {{h_B}} \right|$  in the detection of each symbol. In addition, in order to do marginalization, the detector needs to know the statistics of channel gains beforehand. From equations (25)-(32) in \cite{valenti2011noncoherent}, the detector assumes channel gains   $\left| {{h_A}} \right|$  and $\left| {{h_B}} \right|$   are Rayleigh distributed, and the detector has the knowledge of  $E\left( {{{\left| {{h_A}} \right|}^2}} \right)$ and $E\left( {{{\left| {{h_B}} \right|}^2}} \right)$. This detector serves as a benchmark in this paper.
\item \textbf{KD-MPD}: A detector that applies a channel-gains estimator to estimate channel gains and a `marginalized-phased detector (MPD)' to detect symbols. Specifically, the channel gains estimator is the same as that we described in this section, with KD being used in the first process (i.e., in the first block of Fig.\ref{Fig9}). In addition, MPD makes decision by marginalizing the relative phase between $[0,2\pi )$. The supplement of MPD with the channel gains estimator is referred to as KD-MPD. Specifically, Fig.\ref{Fig8} shows KD-BPD; if the second block of Fig.\ref{Fig8} is replaced by MPD, we have KD-MPD. A special case of this detector has been discussed in Section IV, where channel gains are assumed to be $\left| {{h_A}} \right| = \left| {{h_B}} \right| = 1$ and known at MPD.
\end{enumerate}

In Fig.\ref{Fig10}, the blue lines are PerfPGD, the green lines are KD-BPD, and the red lines are KD. We compare the three detectors under different power-imbalanced channels, where ${\left| {{h_B}} \right|}/{\left| {{h_A}} \right|}$  in Fig.\ref{Fig10} (a), (b), and (c) are 1, 2, and 10 respectively. The channel gains are constant among different simulated packets under each ${\left| {{h_B}} \right|}/{\left| {{h_A}} \right|}$ setting\footnote{For the same detector, the performance is mainly determined by the smaller channel gain between $\left| {{h_A}} \right|$  and $\left| {{h_B}} \right|$. This claim is supported by the following observations: In Fig.\ref{Fig10} (a), (b), and (c), the dashed lines are the cases where $\left| {{h_A}} \right| = 1$  (the smaller channel gain) while $\left| {{h_B}} \right|$  varies from 1 to 10; the solid lines are the cases where $\left| {{h_A}} \right| = 2$   (the smaller channel gain) while $\left| {{h_B}} \right|$  varies from 2 to 20. For the same detector, all the solid lines of the same detector have nearly the same performance among them; so do all the dashed lines of the same detector. Furthermore, there is roughly a 6 dB performance gap between the sold lines and the dashed lines. The results suggest that, for a given detector, it is the smaller channel gain $\left| {{h_A}} \right|$, but not the larger channel gain $\left| {{h_B}} \right|$, that determines the BER performance.}. Fig.\ref{Fig10} shows that KD-BPD has nearly the same performance as the ideal PerfPGD, validating the near-perfect performance of KD-BPD.

KD is a simpler detector compared with KD-BPD since it makes decisions directly without estimating phase and channel gains internally. Fig.\ref{Fig10} shows that KD has nearly 2.8 dB performance gap compared with PerfPGD at ${\rm{BER}} = {10^{ - {\rm{4}}}}$ in different power-imbalanced channels.

Fig.\ref{Fig11} presents the BER under Rayleigh fading channels where the average channel gains for users \emph{A} and \emph{B} are the equal. The instantaneous channel gains for each pair of packets from users \emph{A} and \emph{B}, however, may be power-imbalanced due to fading. In Fig.\ref{Fig11}, the line annotated with circle is KD, the line with squares is MGD, the line with cross markers is KD-MPD, and the line with triangle markers is KD-BPD. KD does not require the explicit values of channel gains and phases, nor the distribution of channel gains and phases. It even does not need to know the noise power. MGD, however, requires that the channel gains are Rayleigh distributed, and also the knowledge of the average channel gains $E\left( {{{\left| {{h_A}} \right|}^2}} \right)$  and $E\left( {{{\left| {{h_B}} \right|}^2}} \right)$. In order to have a comparison with MGD, in Fig.\ref{Fig11}, we assume that the channel gains are Rayleigh distributed and that MGD perfectly knows that $E\left( {{{\left| {{h_A}} \right|}^2}} \right) = E\left( {{{\left| {{h_B}} \right|}^2}} \right) = 1$ (i.e., we gives handicaps to MGD because the other detectors in Fig.\ref{Fig11} do not have such prior knowledge). Even though MGD has this additional knowledge, KD performs better than MGD. Fig.\ref{Fig11} shows that KD has 3.81 dB performance improvement compared with MGD at ${\rm{BER}} = {10^{ - 3}}$.

Unlike BPD, MPD does not estimate the relative phases internally. Instead, MPD just marginalizes over relative phases. Therefore, in general, KD-BPD performs better than KD-MPD. Fig.\ref{Fig11}, for example, shows that KD-BPD performs 1.17 dB better than KD-MPD at ${\rm{BER}} = {10^{ - 3}}$.

\subsection{Implementation Issues}
We discuss two issues related to implementation below:
\begin{enumerate}
\item \textbf{Channel models and channel-gain-and-relative-phase estimation}:
      KD and KD-BPD are general and can be applied to different channel models, including different received powers from users \emph{A} and \emph{B}, different channel fading models, etc. KD does not require channel-gain-and-relative-phase estimation, while KD-BPD estimates the channel-gain-and-relative-phase anew for each pair of new packets from \emph{A} and \emph{B}. As with KD, MGD does not estimate the channel-gain-and-relative-phase; however, it assumes that it knows the channel-gain distribution. Specifically, MGD assumes that the channel gains are Rayleigh distributed, and that $E\left( {{{\left| {{h_A}} \right|}^2}} \right)$  and $E\left( {{{\left| {{h_B}} \right|}^2}} \right)$ are known as \emph{a priori}. Although it can be generalized to other fading channel models, a new detection rule will need to be reconstructed for each fading model. Without this prior information on the channel model, MGD will be far from optimal. That is, if its detection rule under Rayleigh fading is applied to other fading scenarios, the performance will be subpar. Unlike MGD, KD can be applied to any block fading scenarios.\\
      Both KD-BPD and KD-MPD perform channel-gain estimation before BPD and MPD are applied. BPD, however, further estimates the relative phases within its construct, but MPD just marginalizes over all relative phases in its detection rule. As shown in Section IV, the specific phases have significant performance impact and therefore their knowledge is important for superior performance.
\item \textbf{Complexity}: We compute complexity by looking at floating point arithmetic operations and assume that \textbf{a)} addition, subtraction, multiplication, division and square-root between real numbers, each takes one flop\cite{ golub2012matrix }; \textbf{b)} functions such as $\exp\left(  \bullet  \right)$, $\log\left(  \bullet  \right)$, and ${I_0}\left(  \bullet  \right)$  are obtained by table look-up, and thus consume no computation; \textbf{c)} only the dominant term (multiples of \emph{N}) is considered in the complexity computation.\\
    As an example, we show the computation of the complexity of KD here. KD consists of two steps: Assignment step and Update step. The assignment of each symbol involves two subtractions and two multiplications. With \emph{N} symbols in a packet, the Assignment step requires 4\emph{N} flops in total. The Update step involves \emph{N}-2 additions and 2 divisions for a total of \emph{N} flops. Let ${c_1}$  be the number of iterations required for the K-means clustering algorithm to converge. The complexity of KD is therefore  $5{c_1}N$ flops. In general, ${c_1}$   may vary from packet to packet depending on the specific SNR. On average, ${c_1} = 3.54$  in our investigation. Thus, the overall complexity of KD is 17.7\emph{N} flops.\\
    The computation of the complexities of the other detectors are similar. We omit the details and just give the results here. The complexity of MGD is 27\emph{N} flops\footnote{The complexity of MGD is computed based on the closed form of the detector in equation (32) in \cite{valenti2011noncoherent}.}. The complexity of KD-BPD is $\left( {5{c_1}{\rm{ + }}104{c_2}{c_3} + 2} \right)N$  flops, where ${c_2}$  is the quantization levels for the relative phase ${\tilde \theta _n}$  and ${c_3}$  is the quantization levels for $\vartheta $  in BPD. In our simulations,  ${c_2} = 40$ and ${c_3} = 40$, and thus the overall complexity of KD-BPD is 166419.7\emph{N} flops. The complexity of KD-MPD is $\left( {5{c_1} + 59{c_{\rm{4}}} + 41} \right)N$  flops, where  ${c_{\rm{4}}}$ is the quantization levels for the relative phase ${\tilde \theta _n}$  in MPD. In our simulations, ${c_{\rm{4}}} = 40$, and thus the complexity of KD-MPD is 2418.7\emph{N} flops.\\
    In summary, the complexity of KD is smaller than MGD; yet it has better performance than MGD. The complexities of KD-MPD and KD-BPD are greater than those of KD and MGD. Overall, KD-BPD has the highest complexity, but it has the best performance among the four detectors. In terms of order of complexity, all the four detectors have complexities in the order of \emph{N}.
\end{enumerate}

\section{Conclusion}
This paper investigated noncoherent detection for PNC, assuming FSK modulation and short packet transmissions. We found that noncoherent detection in PNC is fundamentally different from noncoherent detection in a single-user point-to-point system: the performance of the noncoherent detector in a single-user point-to-point system is independent of the phase of the received signal; the performance of the noncoherent detector in PNC, however, depends on the relative phase between different transmitters. For good performance of PNC, the noncoherent detector must exploit the knowledge of the relative phase.

Compared with prior work on noncoherent FSK-PNC \cite{sorensen2009physical} \cite{valenti2011noncoherent}, our work is a more comprehensive treatment in two respects: 1) We consider a more general set-up in which neither the channel gains nor the relative phase is known a priori and show that they can be estimated directly from the signal magnitudes given by a simple signal envelope detector. 2) We further show that using the estimated relative phase in noncoherent detection in FSK-PNC can lead to significant performance improvement. In particular, our noncoherent detector for FSK-PNC has nearly the same performance as a fictitious optimal detector that has perfect knowledge of the channel gains and relative phase under general power-imbalanced settings in which different users have different channel gains.

Although this paper focuses on PNC with FSK  modulation, we believe the insight of this paper applies generally to noncoherent detection in other multiuser systems with other modulations. Specifically, our insight is that the relative phase of overlapped signals affects the signal magnitude in multiuser systems, but fortunately the relative phase can be deduced from the magnitudes and this knowledge can be used to improve detection performance.

\appendices
\section{Notations}
The notations in this paper are summarized below:
\begin{basedescript}{\desclabelstyle{\pushlabel}\desclabelwidth{7em}}
\item[$f_u^{\rm{RF}}$\hspace{5.05em}:] The RF frequency of user \emph{u}.
\item[${f_{1,u}}$\hspace{5.2em}:] The first transmitted frequency of user \emph{u}.
\item[${f_{2,u}}$\hspace{5.2em}:] The second transmitted frequency of user \emph{u}.
\item[$\varphi _u^{\rm{RF}}$\hspace{5.05em}:] The initial phase of the RF of user \emph{u}  at the beginning of a packet.
\item[${\varphi _{{h_u}}}$\hspace{5.2em}:] The phase of user \emph{u}'s channel ${h_u}$.
\item[${\varphi _u}$\hspace{5.65em}:] The sum of the channel phase and the RF's initial phase of user \emph{u} (i.e., ${\varphi _u} = \varphi _u^{{\rm{RF}}} + {\varphi _{{h_u}}}$).
\item[$\varphi _{n,u}^{\rm{CFSK}}$\hspace{4.1em}:] User \emph{u}'s phase accumulated over the past \emph{n} symbol periods, assuming the use of continuous-phase FSK modulation at the transmitter. Specifically, $\varphi _{0,u}^{\rm{CFSK}} = 0$; and $\varphi _{n,u}^{\rm{CFSK}} = 2\pi \Delta fT\sum\limits_{i = 0}^{n - 1} {(2{s_{i,u}} - 1)} $  for  $n \ge 1$.
\item[$\tilde f = f_B^{{\rm{RF}}} - f_A^{{\rm{RF}}}$\hspace{0.3em}:] CFO between \emph{A} and \emph{B}.
\item[$\tilde \varphi  = {\varphi _B} - {\varphi _A}$\hspace{1em}:] The initial relative phase between \emph{A} and \emph{B}.
\item[${\tilde \theta _n}$\hspace{5.7em}:] The relative phase between \emph{A} and \emph{B} for symbol \emph{n}. Specifically ${\tilde \theta _n} = 2\pi \tilde fnT + \tilde \varphi  + \varphi _{n,B}^{\rm{CFSK}} - \varphi _{n,A}^{\rm{CFSK}}$.
\end{basedescript}

\section{Further derivation of ${\tilde \theta _n}$}
From \eqref{eq:II-13} in Section III, we know that
\begin{align}
{\tilde \theta _n} = 2n\pi \tilde fT + \tilde \varphi  + \varphi _{n,B}^{{\rm{CFSK}}} - \varphi _{n,A}^{{\rm{CFSK}}} \label{eq:APP_B1}
\end{align}
In \eqref{eq:APP_B1}, for $n \in \left[ {1,N - 1} \right]$, the term $\varphi _{n,B}^{\rm{CFSK}} - \varphi _{n,A}^{\rm{CFSK}}$ can be further expressed as
\begin{align}
\varphi _{n,B}^{\rm{CFSK}} - \varphi _{n,A}^{\rm{CFSK}} = 4\pi \Delta fT\sum\limits_{i = 0}^{n - 1} {\left( {{s_{i,B}} - {s_{i,A}}} \right)}
\end{align}
For noncoherent detection $\Delta f = \frac{1}{{2T}}$ \cite{goldsmith2005wireless}. Thus, we have
\begin{align}
\varphi _{n,B}^{\rm{CFSK}} - \varphi _{n,A}^{\rm{CFSK}} = 2\pi \sum\limits_{i = 0}^{n - 1} {\left( {{s_{i,B}} - {s_{i,A}}} \right)} {\rm{ = }}2\pi {k_n}
\end{align}
where ${k_n}$ is an integer for $n \in \left[ {1,N - 1} \right]$. The second equality follows from the fact that ${s_{n,u}} \in \{ 0,1\}$  for $u \in \left\{ {A,B} \right\}$  and $n = 1 \cdots N - 1$; For $n = 0$,
\begin{align}
\varphi _{0,B}^{\rm{CFSK}} - \varphi _{0,A}^{\rm{CFSK}} = 0{\rm{ = }}2\pi {k_0}
\end{align}
for ${k_0} = 0$. In the end, we get
\begin{align}
{\tilde \theta _n} = 2n\pi \tilde fT + \tilde \varphi  + 2\pi {k_n}
\end{align}
where  ${k_n}$ is an integer for $n \in \left[ {0,N - 1} \right]$. Since two phases with a difference of $2\pi {k_n}$, as in  ${\tilde \theta _n} = 2n\pi \tilde fT + \tilde \varphi {\rm{ + }}2\pi {k_n}$, have the same effect on the magnitude of the received signal, we omit the term $2\pi {k_n}$ and assume
\begin{align}
{\tilde \theta _n} = 2n\pi \tilde fT + \tilde \varphi
\end{align}

\section{Proof of the equations \eqref{eq:III-B1} and \eqref{eq:V-3}}
This appendix shows that for ${\tilde \theta _n} \in \left[ {{\rm{0,2}}\pi } \right]$, in the \emph{q}-th block, $q \in \left[ {1,Q} \right]$,
\begin{equation}
\begin{split}
&\Pr\left( {\left. {{s_n},{{\tilde \theta }_n},\vartheta } \right|{{\bf{R}}_q},\left| {{h_A}} \right|,\left| {{h_B}} \right|} \right)\\
&= {\eta _q}\Pr\left( \vartheta  \right)\Pr\left( {{{\bf{r}}_n}\left| {{s_n},{{\tilde \theta }_n},\left| {{h_A}} \right|,\left| {{h_B}} \right|} \right.} \right)\\
&\times \int {d{{\tilde \theta }_{n - 1}} \cdots d{{\tilde \theta }_{\left( {q - 1} \right)L}}} \prod\limits_{i = (q - 1)L}^{n - 1} {\sum\limits_{{s_i}} {\Pr\left( {{{\bf{r}}_i}\left| {{s_i},{{\tilde \theta }_i},\left| {{h_A}} \right|,\left| {{h_B}} \right|} \right.} \right)\delta \left( {{{\tilde \theta }_i} - {{\left[ {{{\tilde \theta }_{i + 1}} - \vartheta } \right]}_{{\rm{2}}\pi }}} \right)} }\\
&\times \int {d{{\tilde \theta }_{n + 1}} \cdots d{{\tilde \theta }_{qL - 1}}} \prod\limits_{i = n + 1}^{qL - 1} {\sum\limits_{{s_i}} {\Pr\left( {{{\bf{r}}_i}\left| {{s_i},{{\tilde \theta }_i},\left| {{h_A}} \right|,\left| {{h_B}} \right|} \right.} \right)\delta \left( {{{\tilde \theta }_i} - {{\left[ {{{\tilde \theta }_{i - 1}} + \vartheta } \right]}_{{\rm{2}}\pi }}} \right)} }
\end{split}
\end{equation}
where ${{\bf{R}}_q} = \left( {{{\bf{r}}_{\left( {q - 1} \right)L}}, \cdots ,{{\bf{r}}_{qL - 1}}} \right)$;  ${\eta _q}$ is a constant in the \emph{q}-th block; $\vartheta  = {\tilde \theta _n} - {\tilde \theta _{n - 1}} = 2\pi \tilde fT$ is the symbol-to-symbol drift of relative phase induced by a constant CFO $\tilde f$  between users \emph{A} and \emph{B}; $\Pr\left( \vartheta  \right)$  is the distribution of $\vartheta $. We assume that $\vartheta $  is uniformly distributed within $\left[ { - {\rm{0}}{\rm{.02}}\pi ,{\rm{0}}{\rm{.02}}\pi } \right]$; and outside $\left[ { - {\rm{0}}{\rm{.02}}\pi ,{\rm{0}}{\rm{.02}}\pi } \right]$, $\Pr\left( \vartheta  \right){\rm{ = 0}}$.

\begin{proof}
$\Pr\left( {\left. {{s_n},{{\tilde \theta }_n},\vartheta } \right|{{\bf{R}}_q},\left| {{h_A}} \right|,\left| {{h_B}} \right|} \right)$ can be expressed as
follows:
\begin{equation}
\begin{split}
&\Pr\left( {\left. {{s_n},{{\tilde \theta }_n},\vartheta } \right|{{\bf{R}}_q},\left| {{h_A}} \right|,\left| {{h_B}} \right|} \right) \\
&= \int {d{{\tilde \theta }_{\left( {q - 1} \right)L}} \cdots d{{\tilde \theta }_{n - 1}}d{{\tilde \theta }_{n + 1}} \cdots d{{\tilde \theta }_{qL - 1}}}\\
&\times \sum\limits_{{s_{\left( {q - 1} \right)L}}, \cdots ,{s_{n - 1}},{s_{n + 1}}, \cdots ,{s_{qL - 1}}} {\Pr\left( {{s_{\left( {q - 1} \right)L}}, \cdots ,{s_{qL - 1}},{{\tilde \theta }_{\left( {q - 1} \right)L}}, \cdots ,{{\tilde \theta }_{qL - 1}},\vartheta \left| {{{\bf{R}}_q},\left| {{h_A}} \right|,\left| {{h_B}} \right|} \right.} \right)}\\
&= 1/{\Pr\left( {{{\bf{R}}_q},\left| {{h_A}} \right|,\left| {{h_B}} \right|} \right)}\int {d{{\tilde \theta }_{\left( {q - 1} \right)L}} \cdots d{{\tilde \theta }_{n - 1}}d{{\tilde \theta }_{n + 1}} \cdots d{{\tilde \theta }_{qL - 1}}}\\
&\times \sum\limits_{{s_{\left( {q - 1} \right)L}}, \cdots ,{s_{n - 1}},{s_{n + 1}}, \cdots ,{s_{qL - 1}}} {\Pr\left( {{{\bf{R}}_q}\left| \begin{array}{l}
{s_{\left( {q - 1} \right)L}}, \cdots ,{s_{qL - 1}},\\
{{\tilde \theta }_{\left( {q - 1} \right)L}}, \cdots ,{{\tilde \theta }_{qL - 1}},\\
\vartheta ,\left| {{h_A}} \right|,\left| {{h_B}} \right|
\end{array} \right.} \right)\Pr\left( \begin{array}{l}
{s_{\left( {q - 1} \right)L}}, \cdots ,{s_{qL - 1}},\\
{{\tilde \theta }_{\left( {q - 1} \right)L}}, \cdots ,{{\tilde \theta }_{qL - 1}},\\
\vartheta ,\left| {{h_A}} \right|,\left| {{h_B}} \right|
\end{array} \right)}
\end{split}\label{eq:Da}
\end{equation}

In \eqref{eq:Da}, since  $\left( {{s_{\left( {q - 1} \right)L}}, \cdots ,{s_{qL - 1}}} \right)$,  $\left( {{{\tilde \theta }_{\left( {q - 1} \right)L}}, \cdots ,{{\tilde \theta }_{qL - 1}},\vartheta } \right)$,  $\left| {{h_A}} \right|$, and  $\left| {{h_B}} \right|$ are independent of each other, we have
\begin{equation}
\begin{split}
&\Pr\left( {{s_{\left( {q - 1} \right)L}}, \cdots ,{s_{qL - 1}},{{\tilde \theta }_{\left( {q - 1} \right)L}}, \cdots ,{{\tilde \theta }_{qL - 1}},\vartheta ,\left| {{h_A}} \right|,\left| {{h_B}} \right|} \right)\\
&=\Pr\left( {\left| {{h_A}} \right|,\left| {{h_B}} \right|} \right)\Pr\left( {{s_{\left( {q - 1} \right)L}}, \cdots ,{s_{qL - 1}}} \right)\Pr\left( {{{\tilde \theta }_{\left( {q - 1} \right)L}}, \cdots ,{{\tilde \theta }_{qL - 1}},\vartheta } \right)\\
&=\Pr\left( {\left| {{h_A}} \right|,\left| {{h_B}} \right|} \right)\Pr\left( {{s_{\left( {q - 1} \right)L}}, \cdots ,{s_{qL - 1}}} \right)\\
&\times \Pr\left( {{{\tilde \theta }_{\left( {q - 1} \right)L}}, \cdots ,{{\tilde \theta }_{n - 1}},{{\tilde \theta }_{n + 1}}, \cdots ,{{\tilde \theta }_{qL - 1}}\left| {\vartheta ,{{\tilde \theta }_n}} \right.} \right)\Pr( \vartheta,\tilde\theta_n)
\end{split}\label{eq:Dbb}
\end{equation}
Since $\vartheta $ and ${\tilde \theta _n}$ are independent with each other, we have $\Pr\left( {\vartheta ,{{\tilde \theta }_n}} \right) = \Pr\left( \vartheta  \right)\Pr\left( {{{\tilde \theta }_n}} \right)$. In this case, \eqref{eq:Dbb} can be rewritten as
\begin{equation}
\begin{split}
&\Pr\left( {{s_{\left( {q - 1} \right)L}}, \cdots ,{s_{qL - 1}},{{\tilde \theta }_{\left( {q - 1} \right)L}}, \cdots ,{{\tilde \theta }_{qL - 1}},\vartheta ,\left| {{h_A}} \right|,\left| {{h_B}} \right|} \right)\\
&=\Pr\left( {\left| {{h_A}} \right|,\left| {{h_B}} \right|} \right)\Pr\left( {{s_{\left( {q - 1} \right)L}}, \cdots ,{s_{qL - 1}}} \right)\\
&\times \Pr\left( {{{\tilde \theta }_{\left( {q - 1} \right)L}}, \cdots ,{{\tilde \theta }_{n - 1}},{{\tilde \theta }_{n + 1}}, \cdots ,{{\tilde \theta }_{qL - 1}}\left| {\vartheta ,{{\tilde \theta }_n}} \right.} \right)\Pr\left( \vartheta  \right)\Pr\left( {{{\tilde \theta }_n}} \right)
\end{split}\label{eq:Db}
\end{equation}
Note that  ${\tilde \theta _n}$ is assumed to be uniformly distributed between  $\left[ {{\rm{0,2}}\pi } \right]$. In addition, $\Pr\left( \vartheta  \right)$ is the \emph{a priori} distribution of  $\vartheta $. In practice, we should have a rough idea of the range of the CFO  $\tilde f$, and hence  $\vartheta $. The range of CFO investigated in this paper is $\tilde f \in \left[ { - 10{\kern 1pt} {\kern 1pt} {\rm{kHz}},10{\kern 1pt} {\kern 1pt} {\kern 1pt} {\rm{kHz}}} \right]$ (the oscillators in software-defined radio boards typically have CFO smaller than this range), and hence  $\vartheta  \in \left[ { - {\rm{0}}{\rm{.02}}\pi ,{\rm{0}}{\rm{.02}}\pi } \right]$. As a conservative measure, we assume we do not have further information about $\vartheta $  except that it falls within the said range. In particular, we assume  $\vartheta $ is uniformly distributed within $\left[ { - {\rm{0}}{\rm{.02}}\pi ,{\rm{0}}{\rm{.02}}\pi } \right]$, and outside  $\left[ { - {\rm{0}}{\rm{.02}}\pi ,{\rm{0}}{\rm{.02}}\pi } \right]$,  $\Pr\left( \vartheta  \right){\rm{ = 0}}$. In \eqref{eq:Db}, given $\vartheta $  and  ${\tilde \theta _n}$, we can compute the values for  ${\tilde \theta _i} \in \left[ {{\rm{0,2}}\pi } \right]$, $i \in \left[ {\left( {q - 1} \right)L,qL - 1} \right]\backslash n$. Specifically,  ${\tilde \theta _{n - 1}}{\rm{ = }}{\left[ {{{\tilde \theta }_n} - \vartheta } \right]_{2\pi }}$,..., ${\tilde \theta _{\left( {q - 1} \right)L}}{\rm{ = }}{\left[ {{{\tilde \theta }_{\left( {q - 1} \right)L + {\rm{1}}}} - \vartheta } \right]_{2\pi }}$;  ${\tilde \theta _{n + 1}}{\rm{ = }}{[{\tilde \theta _n} + \vartheta ]_{2\pi }}$,..., ${\theta _{qL - 1}}{\rm{ = }}{\left[ {{\theta _{qL - 2}} + \vartheta } \right]_{2\pi }}$, where ${\left[  \bullet  \right]_{2\pi }}{\rm{ = }} \bullet \bmod 2\pi$. In this case, in \eqref{eq:Db},
\begin{equation}
\begin{split}
&\Pr\left( {{{\tilde \theta }_{\left( {q - 1} \right)L}}, \cdots ,{{\tilde \theta }_{n - 1}},{{\tilde \theta }_{n + 1}}, \cdots ,{{\tilde \theta }_{qL - 1}}\left| {\vartheta ,{{\tilde \theta }_n}} \right.} \right)\\
&=\delta \left( {{{\tilde \theta }_{\left( {q - 1} \right)L}} - {{\left[ {{{\tilde \theta }_{\left( {q - 1} \right)L + {\rm{1}}}} - \vartheta } \right]}_{2\pi }}} \right) \cdots \delta \left( {{{\tilde \theta }_{n - 1}} - {{\left[ {{{\tilde \theta }_n} - \vartheta } \right]}_{2\pi }}} \right)\\
&\times \delta \left( {{{\tilde \theta }_{n + 1}} - {{[{{\tilde \theta }_n} + \vartheta ]}_{2\pi }}} \right) \cdots \delta \left( {{\theta _{qL - 1}} - {{\left[ {{\theta _{qL - 2}} + \vartheta } \right]}_{2\pi }}} \right)
\end{split}
\end{equation}

In addition, in \eqref{eq:Da},
\begin{equation}
\begin{split}
&\Pr\left( {{{\bf{R}}_q}\left| {{s_{\left( {q - 1} \right)L}}, \cdots ,{s_{qL - 1}},{{\tilde \theta }_{\left( {q - 1} \right)L}}, \cdots ,{{\tilde \theta }_{qL - 1}},\vartheta ,\left| {{h_A}} \right|,\left| {{h_B}} \right|} \right.} \right)\\
&=\Pr\left( {{{\bf{r}}_{\left( {q - 1} \right)L}}\left| \begin{array}{l}
{{\bf{r}}_{\left( {q - 1} \right)L + 1}}, \cdots ,{{\bf{r}}_{qL - 1}},\\
{s_{\left( {q - 1} \right)L}}, \cdots ,{s_{qL - 1}},\\
{{\tilde \theta }_{\left( {q - 1} \right)L}}, \cdots ,{{\tilde \theta }_{qL - 1}},\\
\vartheta ,\left| {{h_A}} \right|,\left| {{h_B}} \right|
\end{array} \right.} \right)\Pr\left( {{{\bf{r}}_{\left( {q - 1} \right)L + 1}}, \cdots ,{{\bf{r}}_{qL - 1}}\left| \begin{array}{l}
{s_{\left( {q - 1} \right)L}}, \cdots ,{s_{qL - 1}},\\
{{\tilde \theta }_{\left( {q - 1} \right)L}}, \cdots ,{{\tilde \theta }_{qL - 1}},\\
\vartheta ,\left| {{h_A}} \right|,\left| {{h_B}} \right|
\end{array} \right.} \right)\\
&=\Pr\left( {{{\bf{r}}_{\left( {q - 1} \right)L}}\left| \begin{array}{l}
{s_{\left( {q - 1} \right)L}},\\
{{\tilde \theta }_{\left( {q - 1} \right)L}}, \cdots ,{{\tilde \theta }_{qL - 1}},\\
\vartheta ,\left| {{h_A}} \right|,\left| {{h_B}} \right|
\end{array} \right.} \right)\Pr\left( {{{\bf{r}}_{\left( {q - 1} \right)L + 1}}, \cdots ,{{\bf{r}}_{qL - 1}}\left| \begin{array}{l}
{s_{\left( {q - 1} \right)L + 1}}, \cdots ,{s_{qL - 1}},\\
{{\tilde \theta }_{\left( {q - 1} \right)L + {\rm{1}}}}, \cdots ,{{\tilde \theta }_{qL - 1}},\\
\vartheta ,\left| {{h_A}} \right|,\left| {{h_B}} \right|
\end{array} \right.} \right)\\
&=\Pr\left( {{{\bf{r}}_{\left( {q - 1} \right)L}}\left| \begin{array}{l}
{s_{\left( {q - 1} \right)L}},{{\tilde \theta }_{\left( {q - 1} \right)L}},\\
\left| {{h_A}} \right|,\left| {{h_B}} \right|
\end{array} \right.} \right)\Pr\left( {{{\bf{r}}_{\left( {q - 1} \right)L + 1}}, \cdots ,{{\bf{r}}_{qL - 1}}\left| \begin{array}{l}
{s_{\left( {q - 1} \right)L + 1}}, \cdots ,{s_{qL - 1}},\\
{{\tilde \theta }_{\left( {q - 1} \right)L + {\rm{1}}}}, \cdots ,{{\tilde \theta }_{qL - 1}},\\
\vartheta ,\left| {{h_A}} \right|,\left| {{h_B}} \right|
\end{array} \right.} \right)\\
\end{split} \label{eq:De}
\end{equation}
In the above, the second line to the third line follows from the fact that given ${\tilde \theta _{\left( {q - 1} \right)L}}, \cdots ,{\tilde \theta _{qL - 1}}$, $\vartheta$, $\left| {{h_A}} \right|$, and  $\left| {{h_B}} \right|$, ${{\bf{r}}_{\left( {q - 1} \right)L}}$ is independent of $\left( {{{\bf{r}}_{\left( {q - 1} \right)L + 1}}, \cdots ,{{\bf{r}}_{qL - 1}},{s_{\left( {q - 1} \right)L + {\rm{1}}}}, \cdots ,{s_{qL - 1}}} \right)$. The third line to the fourth line follows from the fact that ${{\bf{r}}_{\left( {q - 1} \right)L}}$  only depends on  ${s_{\left( {q - 1} \right)L}}$,  ${\tilde \theta _{\left( {q - 1} \right)L}}$,  $\left| {{h_A}} \right|$, and $\left| {{h_B}} \right|$. Applying the argument in \eqref{eq:De} repeatedly over successive ${{\bf{r}}_i}$, we then get
\begin{equation}
\begin{split}
&\Pr\left( {{{\bf{R}}_q}\left| {{s_{\left( {q - 1} \right)L}}, \cdots ,{s_{qL - 1}},{{\tilde \theta }_{\left( {q - 1} \right)L}}, \cdots ,{{\tilde \theta }_{qL - 1}},\vartheta ,\left| {{h_A}} \right|,\left| {{h_B}} \right|} \right.} \right)\\
&=\prod\limits_{i = (q - 1)L}^{qL - 1} {\Pr\left( {{{\bf{r}}_i}\left| {{s_i},{{\tilde \theta }_i},\left| {{h_A}} \right|,\left| {{h_B}} \right|} \right.} \right)}
\end{split}
\end{equation}

Let ${\eta _q}={\Pr\left( {{s_{\left( {q - 1} \right)L}}, \cdots ,{s_{qL - 1}}} \right)}/{\Pr\left( {\left. {{{\bf{R}}_q}} \right|\left| {{h_A}} \right|,\left| {{h_B}} \right|} \right)}\Pr\left( {{{\tilde \theta }_n}} \right)$, for  ${\tilde \theta _n} \in \left[ {{\rm{0,2}}\pi } \right]$, \eqref{eq:Da} can be expressed as
\begin{equation}
\begin{split}
&\Pr\left( {\left. {{s_n},{{\tilde \theta }_n},\vartheta } \right|{{\bf{R}}_q},\left| {{h_A}} \right|,\left| {{h_B}} \right|} \right)\\
&= {\eta _q}\Pr\left( \vartheta  \right)\Pr\left( {{{\bf{r}}_n}\left| {{s_n},{{\tilde \theta }_n},\left| {{h_A}} \right|,\left| {{h_B}} \right|} \right.} \right)\\
&\times \int {d{{\tilde \theta }_{n - 1}} \cdots d{{\tilde \theta }_{\left( {q - 1} \right)L}}} \prod\limits_{i = (q - 1)L}^{n - 1} {\sum\limits_{{s_i}} {\Pr\left( {{{\bf{r}}_i}\left| {{s_i},{{\tilde \theta }_i},\left| {{h_A}} \right|,\left| {{h_B}} \right|} \right.} \right)\delta \left( {{{\tilde \theta }_i} - {{\left[ {{{\tilde \theta }_{i + 1}} - \vartheta } \right]}_{{\rm{2}}\pi }}} \right)} }\\
&\times \int {d{{\tilde \theta }_{n + 1}} \cdots d{{\tilde \theta }_{qL - 1}}} \prod\limits_{i = n + 1}^{qL - 1} {\sum\limits_{{s_i}} {\Pr\left( {{{\bf{r}}_i}\left| {{s_i},{{\tilde \theta }_i},\left| {{h_A}} \right|,\left| {{h_B}} \right|} \right.} \right)\delta \left( {{{\tilde \theta }_i} - {{\left[ {{{\tilde \theta }_{i - 1}} + \vartheta } \right]}_{{\rm{2}}\pi }}} \right)} }
\end{split} \label{eq:Dc}
\end{equation}
as desired.
\end{proof}

\section{Derivation of PDF $\Pr\left( {{{\bf{r}}_n}\left| {{s_n},{\kern 1pt} {{\tilde \theta }_n}} \right.,\left| {{h_A}} \right|,\left| {{h_B}} \right|} \right)$}
This Appendix derives the expressions of  $\Pr\left( {{{\bf{r}}_n}\left| {{s_n}} \right. = 1,{{\tilde \theta }_n},\left| {{h_A}} \right|,\left| {{h_B}} \right|} \right)$ and\\ $\Pr\left( {{{\bf{r}}_n}\left| {{s_n}} \right.{\rm{ = }}0,{\kern 1pt} {{\tilde \theta }_n},\left| {{h_A}} \right|,\left| {{h_B}} \right|} \right)$ separately, where ${{\bf{r}}_n}{\rm{ = }}\left( {\left| {{r_{n,1}}} \right|,\left| {{r_{n,2}}} \right|} \right)$ is given by
\begin{align}
 {{\bf{r}}_n}=
  \begin{cases}
   \left( {\left| {\left| {{h_A}} \right|{e^{ - j{{{{\tilde \theta }_n}} \mathord{\left/
   {\vphantom {{{{\tilde \theta }_n}} 2}} \right.
   \kern-\nulldelimiterspace} 2}}} + \left| {{h_B}} \right|{e^{j{{{{\tilde \theta }_n}} \mathord{\left/
   {\vphantom {{{{\tilde \theta }_n}} 2}} \right.
   \kern-\nulldelimiterspace} 2}}} + {w_{n,1}}} \right|,\left| {{w_{n,2}}} \right|} \right)& \text{if ${s_{n,A}} = 0,{s_{n,B}} = 0$}\\
   \left( {{\kern 1pt} \left| {{w_{n,1}}} \right|,\left| {\left| {{h_A}} \right|{e^{ - j{{{{\tilde \theta }_n}} \mathord{\left/
   {\vphantom {{{{\tilde \theta }_n}} 2}} \right.
   \kern-\nulldelimiterspace} 2}}} + \left| {{h_B}} \right|{e^{j{{{{\tilde \theta }_n}} \mathord{\left/
   {\vphantom {{{{\tilde \theta }_n}} 2}} \right.
   \kern-\nulldelimiterspace} 2}}} + {w_{n,2}}} \right|} \right)& \text{if ${s_{n,A}} = 1,{s_{n,B}} = 1$}\\
   \left( {\left| {\left| {{h_A}} \right|{e^{ - j{{{{\tilde \theta }_n}} \mathord{\left/
   {\vphantom {{{{\tilde \theta }_n}} 2}} \right.
   \kern-\nulldelimiterspace} 2}}} + {w_{n,1}}} \right|,\left| {\left| {{h_B}} \right|{e^{j{{{{\tilde \theta }_n}} \mathord{\left/
   {\vphantom {{{{\tilde \theta }_n}} 2}} \right.
   \kern-\nulldelimiterspace} 2}}} + {w_{n,2}}} \right|} \right)& \text{if ${s_{n,A}} = 0,{s_{n,B}} = 1$}\\
   \left( {{\kern 1pt} \left| {\left| {{h_B}} \right|{e^{j{{{{\tilde \theta }_n}} \mathord{\left/
   {\vphantom {{{{\tilde \theta }_n}} 2}} \right.
   \kern-\nulldelimiterspace} 2}}} + {w_{n,1}}} \right|,\left| {\left| {{h_A}} \right|{e^{ - j{{{{\tilde \theta }_n}} \mathord{\left/
   {\vphantom {{{{\tilde \theta }_n}} 2}} \right.
   \kern-\nulldelimiterspace} 2}}} + {w_{n,2}}} \right|} \right)& \text{if ${s_{n,A}} = 1,{s_{n,B}} = 0$}
  \end{cases}
\end{align}

We first derive  $\Pr\left( {{{\bf{r}}_n}\left| {{s_n}} \right. = 1,{{\tilde \theta }_n},\left| {{h_A}} \right|,\left| {{h_B}} \right|} \right)$. When \emph{A} and \emph{B} transmit on different frequencies, the magnitudes of the received signal ${{\bf{r}}_n}$  does not depend on the relative phase. Therefore, $\Pr\left( {{{\bf{r}}_n}\left| {{s_n}} \right. = 1,{{\tilde \theta }_n},\left| {{h_A}} \right|,\left| {{h_B}} \right|} \right)=\Pr\left( {{{\bf{r}}_n}\left| {{s_n}} \right. = 1,\left| {{h_A}} \right|,\left| {{h_B}} \right|} \right)$. In addition, when  ${s_{n,A}}{\rm{ = 0}}$ and  ${s_{n,B}} = {\rm{1}}$, the signal magnitudes  $\left| {{r_{n,1}}} \right|$ and $\left| {{r_{n,2}}} \right|$ are Rician-distributed. The conditional PDF of $\left| {{r_{n,1}}} \right|$ is
\begin{equation}
\begin{split}
&\Pr\left( {\left| {{r_{n,1}}} \right|\left| {{s_{n,A}}{\rm{ = 0,}}{s_{n,B}} = {\rm{1}},\left| {{h_A}} \right|,\left| {{h_B}} \right|} \right.} \right)\\
&= {{2\left| {{r_{n,1}}} \right|} \mathord{\left/
 {\vphantom {{2\left| {{r_{n,1}}} \right|} {{N_0}}}} \right.
 \kern-\nulldelimiterspace} {{N_0}}}\exp \left\{ { - {{\left( {{{\left| {{r_{n,1}}} \right|}^2} + {{\left| {{h_A}} \right|}^{\rm{2}}}} \right)} \mathord{\left/
 {\vphantom {{\left( {{{\left| {{r_{n,1}}} \right|}^2} + {{\left| {{h_A}} \right|}^{\rm{2}}}} \right)} {{N_0}}}} \right.
 \kern-\nulldelimiterspace} {{N_0}}}} \right\}{I_0}\left( {{{2\left| {{h_A}} \right|\left| {{r_{n,1}}} \right|} \mathord{\left/
 {\vphantom {{2\left| {{h_A}} \right|\left| {{r_{n,1}}} \right|} {{N_0}}}} \right.
 \kern-\nulldelimiterspace} {{N_0}}}} \right)
\end{split}
\end{equation}
where ${I_0}\left( x \right)$ is the modified Bessel function of the first kind of order zero. The
conditional PDF of $\left| {{r_{n,{\rm{2}}}}} \right|$ is
\begin{equation}
\begin{split}
&\Pr\left( {\left| {{r_{n,{\rm{2}}}}} \right|\left| {{s_{n,A}}{\rm{ = 0,}}{s_{n,B}} = {\rm{1}},\left| {{h_A}} \right|,\left| {{h_B}} \right|} \right.} \right)\\
&={{2\left| {{r_{n,{\rm{2}}}}} \right|} \mathord{\left/
 {\vphantom {{2\left| {{r_{n,{\rm{2}}}}} \right|} {{N_0}}}} \right.
 \kern-\nulldelimiterspace} {{N_0}}}\exp \left\{ { - {{\left( {{{\left| {{r_{n,{\rm{2}}}}} \right|}^2} + {{\left| {{h_B}} \right|}^{\rm{2}}}} \right)} \mathord{\left/
 {\vphantom {{\left( {{{\left| {{r_{n,{\rm{2}}}}} \right|}^2} + {{\left| {{h_B}} \right|}^{\rm{2}}}} \right)} {{N_0}}}} \right.
 \kern-\nulldelimiterspace} {{N_0}}}} \right\}{I_0}\left( {{{2\left| {{h_B}} \right|\left| {{r_{n,{\rm{2}}}}} \right|} \mathord{\left/
 {\vphantom {{2\left| {{h_B}} \right|\left| {{r_{n,{\rm{2}}}}} \right|} {{N_0}}}} \right.
 \kern-\nulldelimiterspace} {{N_0}}}} \right)
\end{split}
\end{equation}
Since the signal magnitudes on two different frequencies are independent of each other, we have
\begin{equation}
\begin{split}
&\Pr\left( {\left( {\left| {{r_{n,1}}} \right|,\left| {{r_{n,2}}} \right|} \right)\left| {{s_{n,A}}{\rm{ = 0,}}{s_{n,B}}} \right. = {\rm{1}},\left| {{h_A}} \right|,\left| {{h_B}} \right|} \right)\\
&=\Pr\left( {\left| {{r_{n,1}}} \right|\left| {{s_{n,A}}{\rm{ = 0,}}{s_{n,B}}} \right. = {\rm{1}},\left| {{h_A}} \right|,\left| {{h_B}} \right|} \right)\Pr\left( {\left| {{r_{n,{\rm{2}}}}} \right|\left| {{s_{n,A}}{\rm{ = 0,}}{s_{n,B}}} \right. = {\rm{1}},\left| {{h_A}} \right|,\left| {{h_B}} \right|} \right)\\
&= {{{\rm{4}}\left| {{r_{n,1}}} \right|\left| {{r_{n,{\rm{2}}}}} \right|} \mathord{\left/
 {\vphantom {{{\rm{4}}\left| {{r_{n,1}}} \right|\left| {{r_{n,{\rm{2}}}}} \right|} {N_{\rm{0}}^{\rm{2}}}}} \right.
 \kern-\nulldelimiterspace} {N_{\rm{0}}^{\rm{2}}}}\exp \left\{ { - {{\left( {{{\left| {{r_{n,1}}} \right|}^2}{\rm{ + }}{{\left| {{r_{n,{\rm{2}}}}} \right|}^2} + {{\left| {{h_{\rm{A}}}} \right|}^{\rm{2}}} + {{\left| {{h_B}} \right|}^{\rm{2}}}} \right)} \mathord{\left/
 {\vphantom {{\left( {{{\left| {{r_{n,1}}} \right|}^2}{\rm{ + }}{{\left| {{r_{n,{\rm{2}}}}} \right|}^2} + {{\left| {{h_{\rm{A}}}} \right|}^{\rm{2}}} + {{\left| {{h_B}} \right|}^{\rm{2}}}} \right)} {{N_{\rm{0}}}}}} \right.
 \kern-\nulldelimiterspace} {{N_{\rm{0}}}}}} \right\}\\
&\times {I_0}\left( {{{2\left| {{h_{\rm{A}}}} \right|\left| {{r_{n,1}}} \right|} \mathord{\left/
 {\vphantom {{2\left| {{h_{\rm{A}}}} \right|\left| {{r_{n,1}}} \right|} {{N_{\rm{0}}}}}} \right.
 \kern-\nulldelimiterspace} {{N_{\rm{0}}}}}} \right){I_0}\left( {{{2\left| {{h_B}} \right|\left| {{r_{n,{\rm{2}}}}} \right|} \mathord{\left/
 {\vphantom {{2\left| {{h_B}} \right|\left| {{r_{n,{\rm{2}}}}} \right|} {{N_{\rm{0}}}}}} \right.
 \kern-\nulldelimiterspace} {{N_{\rm{0}}}}}} \right)
\end{split}
\end{equation}
Similarly,
\begin{equation}
\begin{split}
&\Pr\left( {\left( {\left| {{r_{n,1}}} \right|,\left| {{r_{n,2}}} \right|} \right)\left| {{s_{n,A}}{\rm{ = 1,}}{s_{n,B}}} \right. = 0,\left| {{h_A}} \right|,\left| {{h_B}} \right|} \right)\\
&= {{{\rm{4}}\left| {{r_{n,1}}} \right|\left| {{r_{n,{\rm{2}}}}} \right|} \mathord{\left/
 {\vphantom {{{\rm{4}}\left| {{r_{n,1}}} \right|\left| {{r_{n,{\rm{2}}}}} \right|} {N_{\rm{0}}^{\rm{2}}}}} \right.
 \kern-\nulldelimiterspace} {N_{\rm{0}}^{\rm{2}}}}\exp \left\{ { - {{\left( {{{\left| {{r_{n,1}}} \right|}^2}{\rm{ + }}{{\left| {{r_{n,{\rm{2}}}}} \right|}^2} + {{\left| {{h_{\rm{A}}}} \right|}^{\rm{2}}} + {{\left| {{h_B}} \right|}^{\rm{2}}}} \right)} \mathord{\left/
 {\vphantom {{\left( {{{\left| {{r_{n,1}}} \right|}^2}{\rm{ + }}{{\left| {{r_{n,{\rm{2}}}}} \right|}^2} + {{\left| {{h_{\rm{A}}}} \right|}^{\rm{2}}} + {{\left| {{h_B}} \right|}^{\rm{2}}}} \right)} {{N_{\rm{0}}}}}} \right.
 \kern-\nulldelimiterspace} {{N_{\rm{0}}}}}} \right\}\\
&\times {I_0}\left( {{{2\left| {{h_B}} \right|\left| {{r_{n,1}}} \right|} \mathord{\left/
 {\vphantom {{2\left| {{h_B}} \right|\left| {{r_{n,1}}} \right|} {{N_{\rm{0}}}}}} \right.
 \kern-\nulldelimiterspace} {{N_{\rm{0}}}}}} \right){I_0}\left( {{{2\left| {{h_{\rm{A}}}} \right|\left| {{r_{n,{\rm{2}}}}} \right|} \mathord{\left/
 {\vphantom {{2\left| {{h_{\rm{A}}}} \right|\left| {{r_{n,{\rm{2}}}}} \right|} {{N_{\rm{0}}}}}} \right.
 \kern-\nulldelimiterspace} {{N_{\rm{0}}}}}} \right)
\end{split}
\end{equation}
When two users transmit on different frequencies, the conditional PDF is
\begin{equation}
\begin{split}
&\Pr\left( {{{\bf{r}}_n}\left| {{s_n}} \right. = 1,{{\tilde \theta }_n},\left| {{h_A}} \right|,\left| {{h_B}} \right|} \right)\\
&=\frac{{\rm{1}}}{{\rm{2}}}\Pr\left( {{{\bf{r}}_n}\left| {{s_{n,A}}{\rm{ = 0,}}{s_{n,B}}} \right. = {\rm{1}},\left| {{h_A}} \right|,\left| {{h_B}} \right|} \right)\\
&+\frac{{\rm{1}}}{{\rm{2}}}\Pr\left( {{{\bf{r}}_n}\left| {{s_{n,A}}{\rm{ = 1,}}{s_{n,B}}} \right. = {\rm{0}},\left| {{h_A}} \right|,\left| {{h_B}} \right|} \right)
\end{split}
\end{equation}

Next, we derive  ${\rm{P}}\left( {{{\bf{r}}_n}\left| {{s_n} = 0,{{\tilde \theta }_n},\left| {{h_A}} \right|,\left| {{h_B}} \right|} \right.} \right)$. When  ${s_{n,A}}{\rm{ = 0}}$ and ${s_{n,B}} = {\rm{0}}$, $\left| {{r_{n,1}}} \right|$ is Rician-distributed. The conditional PDF of  $\left| {{r_{n,1}}} \right|$ is
\begin{equation}
\begin{split}
&\Pr\left( {\left| {{r_{n,1}}} \right|\left| {{s_{n,A}}{\rm{ = 0,}}{s_{n,B}}} \right. = {\rm{0,}}{{\tilde \theta }_n},\left| {{h_A}} \right|,\left| {{h_B}} \right|} \right)\\
&= {{2\left| {{r_{n,1}}} \right|} \mathord{\left/
 {\vphantom {{2\left| {{r_{n,1}}} \right|} {{N_0}}}} \right.
 \kern-\nulldelimiterspace} {{N_0}}}\exp \left\{ { - {{\left( {{{\left| {{r_{n,1}}} \right|}^2} + {{\left| {{h_A}} \right|}^{\rm{2}}}{\rm{ + }}{{\left| {{h_B}} \right|}^{\rm{2}}}{\rm{ + 2}}\left| {{h_A}} \right|\left| {{h_B}} \right|{\rm{cos}}\left( {{{\tilde \theta }_n}} \right)} \right)} \mathord{\left/
 {\vphantom {{\left( {{{\left| {{r_{n,1}}} \right|}^2} + {{\left| {{h_A}} \right|}^{\rm{2}}}{\rm{ + }}{{\left| {{h_B}} \right|}^{\rm{2}}}{\rm{ + 2}}\left| {{h_A}} \right|\left| {{h_B}} \right|{\rm{cos}}\left( {{{\tilde \theta }_n}} \right)} \right)} {{N_0}}}} \right.
 \kern-\nulldelimiterspace} {{N_0}}}} \right\}\\
&\times {I_0}\left( {{{\left( {2\left| {{r_{n,1}}} \right|\sqrt {{{\left| {{h_A}} \right|}^{\rm{2}}}{\rm{ + }}{{\left| {{h_B}} \right|}^{\rm{2}}}{\rm{ + 2}}\left| {{h_A}} \right|\left| {{h_B}} \right|{\rm{cos}}\left( {{{\tilde \theta }_n}} \right)} } \right)} \mathord{\left/
 {\vphantom {{\left( {2\left| {{r_{n,1}}} \right|\sqrt {{{\left| {{h_A}} \right|}^{\rm{2}}}{\rm{ + }}{{\left| {{h_B}} \right|}^{\rm{2}}}{\rm{ + 2}}\left| {{h_A}} \right|\left| {{h_B}} \right|{\rm{cos}}\left( {{{\tilde \theta }_n}} \right)} } \right)} {{N_0}}}} \right.
 \kern-\nulldelimiterspace} {{N_0}}}} \right)
\end{split}
\end{equation}
Since  $\left| {{r_{n,{\rm{2}}}}} \right|$ contains only noise, it is Rayleigh distributed. The PDF is
\begin{align}
\Pr\left( {\left| {{r_{n,{\rm{2}}}}} \right|\left| {{s_{n,A}}{\rm{ = 0,}}{s_{n,B}}} \right. = {\rm{0,}}{{\tilde \theta }_n},\left| {{h_A}} \right|,\left| {{h_B}} \right|} \right) = {{2\left| {{r_{n,{\rm{2}}}}} \right|} \mathord{\left/
 {\vphantom {{2\left| {{r_{n,{\rm{2}}}}} \right|} {{N_0}}}} \right.
 \kern-\nulldelimiterspace} {{N_0}}}\exp \left\{ { - {{{{\left| {{r_{n,{\rm{2}}}}} \right|}^2}} \mathord{\left/
 {\vphantom {{{{\left| {{r_{n,{\rm{2}}}}} \right|}^2}} {{N_{\rm{0}}}}}} \right.
 \kern-\nulldelimiterspace} {{N_{\rm{0}}}}}} \right\}
\end{align}
Therefore,
\begin{equation}
\begin{split}
&\Pr\left( {\left( {\left| {{r_{n,1}}} \right|,\left| {{r_{n,2}}} \right|} \right)\left| {{s_{n,A}}{\rm{ = 0,}}{s_{n,B}}} \right. = 0,{{\tilde \theta }_n},\left| {{h_A}} \right|,\left| {{h_B}} \right|} \right)\\
&=\Pr\left( {\left| {{r_{n,1}}} \right|\left| {{s_{n,A}}{\rm{ = 0,}}{s_{n,B}}} \right. = 0,{{\tilde \theta }_n},\left| {{h_A}} \right|,\left| {{h_B}} \right|} \right)\Pr\left( {\left| {{r_{n,{\rm{2}}}}} \right|\left| {{s_{n,A}}{\rm{ = 0,}}{s_{n,B}}} \right. = 0,{{\tilde \theta }_n},\left| {{h_A}} \right|,\left| {{h_B}} \right|} \right)\\
&= {{{\rm{4}}\left| {{r_{n,1}}} \right|\left| {{r_{n,{\rm{2}}}}} \right|} \mathord{\left/
 {\vphantom {{{\rm{4}}\left| {{r_{n,1}}} \right|\left| {{r_{n,{\rm{2}}}}} \right|} {N_{\rm{0}}^{\rm{2}}}}} \right.
 \kern-\nulldelimiterspace} {N_{\rm{0}}^{\rm{2}}}}\exp \left\{ { - {{\left( {{{\left| {{r_{n,1}}} \right|}^2}{\rm{ + }}{{\left| {{r_{n,{\rm{2}}}}} \right|}^2} + {{\left| {{h_A}} \right|}^{\rm{2}}}{\rm{ + }}{{\left| {{h_B}} \right|}^{\rm{2}}}{\rm{ + 2}}\left| {{h_A}} \right|\left| {{h_B}} \right|{\rm{cos}}\left( {{{\tilde \theta }_n}} \right)} \right)} \mathord{\left/
 {\vphantom {{\left( {{{\left| {{r_{n,1}}} \right|}^2}{\rm{ + }}{{\left| {{r_{n,{\rm{2}}}}} \right|}^2} + {{\left| {{h_A}} \right|}^{\rm{2}}}{\rm{ + }}{{\left| {{h_B}} \right|}^{\rm{2}}}{\rm{ + 2}}\left| {{h_A}} \right|\left| {{h_B}} \right|{\rm{cos}}\left( {{{\tilde \theta }_n}} \right)} \right)} {{N_{\rm{0}}}}}} \right.
 \kern-\nulldelimiterspace} {{N_{\rm{0}}}}}} \right\}\\
&\times {I_0}\left( {{{\left( {2\left| {{r_{n,1}}} \right|\sqrt {{{\left| {{h_A}} \right|}^{\rm{2}}}{\rm{ + }}{{\left| {{h_B}} \right|}^{\rm{2}}}{\rm{ + 2}}\left| {{h_A}} \right|\left| {{h_B}} \right|{\rm{cos}}\left( {{{\tilde \theta }_n}} \right)} } \right)} \mathord{\left/
 {\vphantom {{\left( {2\left| {{r_{n,1}}} \right|\sqrt {{{\left| {{h_A}} \right|}^{\rm{2}}}{\rm{ + }}{{\left| {{h_B}} \right|}^{\rm{2}}}{\rm{ + 2}}\left| {{h_A}} \right|\left| {{h_B}} \right|{\rm{cos}}\left( {{{\tilde \theta }_n}} \right)} } \right)} {{N_{\rm{0}}}}}} \right.
 \kern-\nulldelimiterspace} {{N_{\rm{0}}}}}} \right)
\end{split}
\end{equation}
Similarly,
\begin{equation}
\begin{split}
&\Pr\left( {\left( {\left| {{r_{n,1}}} \right|,\left| {{r_{n,2}}} \right|} \right)\left| {{s_{n,A}}{\rm{ = 1,}}{s_{n,B}}} \right. = {\rm{1,}}{{\tilde \theta }_n},\left| {{h_A}} \right|,\left| {{h_B}} \right|} \right)\\
&= {{{\rm{4}}\left| {{r_{n,1}}} \right|\left| {{r_{n,{\rm{2}}}}} \right|} \mathord{\left/
 {\vphantom {{{\rm{4}}\left| {{r_{n,1}}} \right|\left| {{r_{n,{\rm{2}}}}} \right|} {N_{\rm{0}}^{\rm{2}}}}} \right.
 \kern-\nulldelimiterspace} {N_{\rm{0}}^{\rm{2}}}}\exp \left\{ { - {{\left( {{{\left| {{r_{n,1}}} \right|}^2}{\rm{ + }}{{\left| {{r_{n,{\rm{2}}}}} \right|}^2} + {{\left| {{h_A}} \right|}^{\rm{2}}}{\rm{ + }}{{\left| {{h_B}} \right|}^{\rm{2}}}{\rm{ + 2}}\left| {{h_A}} \right|\left| {{h_B}} \right|{\rm{cos}}\left( {{{\tilde \theta }_n}} \right)} \right)} \mathord{\left/
 {\vphantom {{\left( {{{\left| {{r_{n,1}}} \right|}^2}{\rm{ + }}{{\left| {{r_{n,{\rm{2}}}}} \right|}^2} + {{\left| {{h_A}} \right|}^{\rm{2}}}{\rm{ + }}{{\left| {{h_B}} \right|}^{\rm{2}}}{\rm{ + 2}}\left| {{h_A}} \right|\left| {{h_B}} \right|{\rm{cos}}\left( {{{\tilde \theta }_n}} \right)} \right)} {{N_{\rm{0}}}}}} \right.
 \kern-\nulldelimiterspace} {{N_{\rm{0}}}}}} \right\}\\
&\times {I_0}\left( {{{2\left| {{r_{n,2}}} \right|\sqrt {{{\left| {{h_A}} \right|}^{\rm{2}}}{\rm{ + }}{{\left| {{h_B}} \right|}^{\rm{2}}}{\rm{ + 2}}\left| {{h_A}} \right|\left| {{h_B}} \right|{\rm{cos}}\left( {{{\tilde \theta }_n}} \right)} } \mathord{\left/
 {\vphantom {{2\left| {{r_{n,2}}} \right|\sqrt {{{\left| {{h_A}} \right|}^{\rm{2}}}{\rm{ + }}{{\left| {{h_B}} \right|}^{\rm{2}}}{\rm{ + 2}}\left| {{h_A}} \right|\left| {{h_B}} \right|{\rm{cos}}\left( {{{\tilde \theta }_n}} \right)} } {{N_{\rm{0}}}}}} \right.
 \kern-\nulldelimiterspace} {{N_{\rm{0}}}}}} \right)
\end{split}
\end{equation}
Again, overall we have
\begin{equation}
\begin{split}
&\Pr\left( {{{\bf{r}}_n}\left| {{s_n} = 0,{{\tilde \theta }_n},\left| {{h_A}} \right|,\left| {{h_B}} \right|} \right.} \right)\\
&=\frac{{\rm{1}}}{{\rm{2}}}{\rm{P}}\left( {{{\bf{r}}_n}\left| {{s_{n,A}}{\rm{ = 0,}}{s_{n,B}}} \right. = 0,{\kern 1pt} {{\tilde \theta }_n},\left| {{h_A}} \right|,\left| {{h_B}} \right|} \right)\\
&+ \frac{{\rm{1}}}{{\rm{2}}}{\rm{P}}\left( {{{\bf{r}}_n}\left| {{s_{n,A}}{\rm{ = 1,}}{s_{n,B}}} \right. = 1,{\kern 1pt} {{\tilde \theta }_n},\left| {{h_A}} \right|,\left| {{h_B}} \right|} \right)
\end{split}
\end{equation}

\bibliographystyle{IEEEtran}
\bibliography{references}

\begin{thebibliography}{10}
\providecommand{\url}[1]{#1}
\csname url@samestyle\endcsname
\providecommand{\newblock}{\relax}
\providecommand{\bibinfo}[2]{#2}
\providecommand{\BIBentrySTDinterwordspacing}{\spaceskip=0pt\relax}
\providecommand{\BIBentryALTinterwordstretchfactor}{4}
\providecommand{\BIBentryALTinterwordspacing}{\spaceskip=\fontdimen2\font plus
\BIBentryALTinterwordstretchfactor\fontdimen3\font minus
  \fontdimen4\font\relax}
\providecommand{\BIBforeignlanguage}[2]{{%
\expandafter\ifx\csname l@#1\endcsname\relax
\typeout{** WARNING: IEEEtran.bst: No hyphenation pattern has been}%
\typeout{** loaded for the language `#1'. Using the pattern for}%
\typeout{** the default language instead.}%
\else
\language=\csname l@#1\endcsname
\fi
#2}}
\providecommand{\BIBdecl}{\relax}
\BIBdecl

\bibitem{zhang2006hot}
S.~Zhang, S.~C. Liew, and P.~P. Lam, ``Hot topic: Physical-layer network
  coding,'' in \emph{Proceedings of the 12th annual international conference on
  Mobile computing and networking}.\hskip 1em plus 0.5em minus 0.4em\relax ACM,
  2006, pp. 358--365.

\bibitem{popovski2007physical}
P.~Popovski and H.~Yomo, ``Physical network coding in two-way wireless relay
  channels,'' in \emph{Communications, 2007. ICC'07. IEEE International
  Conference on}.\hskip 1em plus 0.5em minus 0.4em\relax IEEE, 2007, pp.
  707--712.

\bibitem{nazer2011compute}
B.~Nazer and M.~Gastpar, ``Compute-and-forward: Harnessing interference through
  structured codes,'' \emph{IEEE Transactions on Information Theory}, vol.~57,
  no.~10, pp. 6463--6486, 2011.

\bibitem{goldsmith2005wireless}
A.~Goldsmith, \emph{Wireless communications}.\hskip 1em plus 0.5em minus
  0.4em\relax Cambridge university press, 2005.

\bibitem{wang2014cellular}
C.-X. Wang, F.~Haider, X.~Gao, X.-H. You, Y.~Yang, D.~Yuan, H.~Aggoune,
  H.~Haas, S.~Fletcher, and E.~Hepsaydir, ``Cellular architecture and key
  technologies for {5G} wireless communication networks,'' \emph{IEEE
  Communications Magazine}, vol.~52, no.~2, pp. 122--130, 2014.

\bibitem{liew2013physical}
S.~C. Liew, S.~Zhang, and L.~Lu, ``Physical-layer network coding: Tutorial,
  survey, and beyond,'' \emph{Physical Communication}, vol.~6, pp. 4--42, 2013.

\bibitem{ProakisJohnG2008Dc}
J.~G. Proakis, \emph{\BIBforeignlanguage{eng}{Digital communications}},
  5th~ed.\hskip 1em plus 0.5em minus 0.4em\relax Boston: McGraw-Hill, 2008.

\bibitem{CouchLeonW2013Daac}
L.~W. Couch, \emph{\BIBforeignlanguage{eng}{Digital and analog communication
  systems}}, 8th~ed.\hskip 1em plus 0.5em minus 0.4em\relax Upper Saddle River,
  N.J.: Pearson, 2013.

\bibitem{yu2016physical}
Q.-Y. Yu, D.-Y. Zhang, H.-H. Chen, and W.-X. Meng, ``Physical-layer network
  coding systems with {MFSK} modulation,'' \emph{IEEE Transactions on Vehicular
  Technology}, vol.~65, no.~1, pp. 204--213, 2016.

\bibitem{sorensen2009physical}
J.~H. Sorensen, R.~Krigslund, P.~Popovski, T.~K. Akino, and T.~Larsen,
  ``Physical layer network coding for {FSK} systems,'' \emph{IEEE
  Communications Letters}, vol.~13, no.~8, pp. 597--599, 2009.

\bibitem{valenti2011noncoherent}
M.~C. Valenti, D.~Torrieri, and T.~Ferrett, ``Noncoherent physical-layer
  network coding with {FSK} modulation: Relay receiver design issues,''
  \emph{IEEE Transactions on Communications}, vol.~59, no.~9, pp. 2595--2604,
  2011.

\bibitem{wang2018dcap}
T.~Wang, Q.~Yang, K.~Tan, J.~Zhang, S.~C. Liew, and S.~Zhang, ``{DCAP}:
  Improving the capacity of {WiFi} networks with distributed cooperative access
  points,'' \emph{IEEE Transactions on Mobile Computing}, vol.~17, no.~2, pp.
  320--333, 2018.

\bibitem{PearlJudea1988Prii}
J.~Pearl, \emph{\BIBforeignlanguage{eng}{Probabilistic reasoning in intelligent
  systems : networks of plausible inference}}, ser. Morgan Kaufmann series in
  representation and reasoning.\hskip 1em plus 0.5em minus 0.4em\relax San
  Mateo, Calif.: Morgan Kaufmann Publishers, 1988.

\bibitem{lu2012asynchronous}
L.~Lu and S.~C. Liew, ``Asynchronous physical-layer network coding,''
  \emph{IEEE Transactions on Wireless Communications}, vol.~11, no.~2, pp.
  819--831, 2012.

\bibitem{shao2017asynchronous}
Y.~Shao, S.~C. Liew, and L.~Lu, ``Asynchronous physical-layer network coding:
  Symbol misalignment estimation and its effect on decoding,'' \emph{IEEE
  Transactions on Wireless Communications}, vol.~16, no.~10, pp. 6881--6894,
  2017.

\bibitem{mackay2003information}
D.~J. MacKay, \emph{Information theory, inference and learning
  algorithms}.\hskip 1em plus 0.5em minus 0.4em\relax Cambridge university
  press, 2003.

\bibitem{golub2012matrix}
G.~H. Golub and C.~F. Van~Loan, \emph{Matrix computations}.\hskip 1em plus
  0.5em minus 0.4em\relax JHU Press, 2012, vol.~3.

\end{thebibliography}
\end{document}